\documentclass[useAMS,usenatbib]{mn2e}
\usepackage{graphicx}
\usepackage{color}
\usepackage{amsmath}
\usepackage{lscape} 
\usepackage{times} 

\title[The incidence of obscuration in AGN]{The incidence of
  obscuration in active galactic nuclei}

\author[Merloni et al.]
{\parbox{\textwidth}{A. Merloni$^{1}$\thanks{E-mail: \texttt{am@mpe.mpg.de (MPE)}},
A. Bongiorno$^{2}$,
M. Brusa$^{1,3,4}$,
K. Iwasawa$^{5}$
V. Mainieri$^{6}$,
B. Magnelli$^{7}$, 
M. Salvato$^{1}$,
S. Berta$^{1}$, 
N. Cappelluti$^{4}$,
A. Comastri$^{4}$,
F. Fiore$^{2}$,
R. Gilli$^{4}$, 
A. Koekemoer$^{8}$,
E. Le Floc'h$^{9}$,
E. Lusso$^{10}$, 
D. Lutz$^{1}$, 
T. Miyaji$^{11,12}$,
F. Pozzi$^{3}$,
L. Riguccini$^{13}$,
D.J. Rosario$^{1}$, 
J. Silverman$^{14}$,
M. Symeonidis$^{15,16}$,
E. Treister$^{17}$,
C. Vignali$^{3}$ and 
G. Zamorani$^{4}$
}\vspace{0.4cm}\\
\parbox{\textwidth}{$^{1}$Max-Planck-Institut f\"ur extraterrestrische
  Physik (MPE), Giessenbachstrasse 1, D-85748, Garching bei M\"unchen,
  Germany\\ 
$^{2}$INAF-Osservatorio Astronomico di Roma, Via di Frascati 33, 00040, Monteporzio Catone, Rome, Italy\\
$^{3}$Dipartimento di Astronomia, Universit{\'a} di Bologna, Via
Ranzani 1, 40127, Bologna, Italy\\
$^{4}$INAF-Osservatorio Astronomico di Bologna, Via Ranzani 1, 40127, Bologna, Italy\\
$^{5}$ICREA and Institut de Ci\`encies del Cosmos (ICC), Universitat de Barcelona (IEEC-UB), Mart\'i i Franqu\`es, 1, 08028 Barcelona, Spain\\
$^{6}$European Southern Observatory, K. Schwarzschildstr. 1, 85741, Garching, Germany\\
$^{7}$Argelander-Institut f{\"u}r Astronomie, Universit{\"a}t Bonn, Auf dem H{\"u}gel 71, 53121 Bonn, Germany\\ 
$^{8}$Space Telescope Science Institute, 3700 San Martin Drive, Baltimore MD 21218, U.S.A. \\
$^{9}$Laboratoire AIM-Paris-Saclay, CEA/DSM/Irfu, CNRS, Universit{\'e} Paris Diderot, Saclay, pt courrier 131, 91191 Gif-sur-Yvette,
France\\
$^{10}$Max Planck Institute for Astronomy, K{\"o}nigstuhl 17, 69117 Heidelberg, Germany\\
$^{11}$Instituto de Astronomia, UNAM, Ensenada, Baja California, Mexico (PO Box 439027, San
Diego, CA 92143-9027, USA)\\
$^{12}$ University of California, San Diego, Center for Astrophysics and Space Sciences, 9500 Gilman Drive, La Jolla, CA 92093-0424,
USA\\
$^{13}$ NASA Ames, Moffett field, CA 94035, USA\\
$^{14}$ Kavli Institute for the Physics and Mathematics of the Universe, The University of Tokyo, 5-1-5 Kashiwanoha, Kashiwashi, Chiba, 277-8583, Japan\\
$^{15}$ University of Sussex, Department of Physics and Astronomy, Falmer, Brighton BN1 9QH, Sussex, UK\\
$^{16}$ Mullard Space Science Laboratory, University College London, Holmbury St. Mary, Dorking, Surrey RH5 6NT, UK\\
$^{17}$ Departamento de Astronomıa, Universidad de Concepcion, Casilla 160-C, Concepcion, Chile\\
}}

\begin{document}

\date{}

\maketitle

\label{firstpage}

\vspace{-0.5cm}
\begin{abstract}
We study the incidence of nuclear obscuration on a complete sample of
1310 AGN selected on the basis of their rest-frame 2--10 keV X-ray flux from
the XMM-COSMOS survey, in the redshift range $0.3<z<3.5$. 
We classify the AGN as obscured or un-obscured on the basis of either
the optical spectral properties and the overall SED or the shape of the X-ray spectrum. 
The two classifications agree in about 70\% of the objects, and the remaining 30\% 
can be further subdivided into two distinct
classes: at low luminosities X-ray un-obscured AGN do not always show signs of broad lines or
blue/UV continuum emission in their optical spectra, most likely due to galaxy dilution effects; at high 
luminosities broad line AGN may have absorbed X-ray
spectra, which hints at
an increased incidence of small-scale (sub-parsec) dust-free obscuration. 
We confirm that the fraction of obscured AGN
is a decreasing function of the intrinsic X-ray luminosity, while the
incidence of absorption shows significant evolution only for the most luminous AGN,
which appear to be more commonly obscured at higher redshift. We find no significant difference between 
the mean stellar masses and star formation rates of obscured and un-obscured AGN hosts.
We conclude that the physical state of the medium responsible for
obscuration in AGN is complex, and mainly determined by
the radiation environment (nuclear luminosity) in a small region enclosed within the gravitational
sphere of influence of the central black
hole, but is largely insensitive to the wider scale galactic conditions. 
\end{abstract}


\section{Introduction}

The observational appearance of an Active Galactic Nucleus (AGN) is not only determined by its
intrinsic emission properties, but also by the nature, amount, dynamical and kinematic state
of any intervening material along the line of sight. Intrinsic obscuration does indeed play
 a fundamental role for our understanding of the overall properties of AGN.
Large, systematic studies of their spectral energy distribution
clearly show that the intrinsic shape of the X-ray continuum can
be characterized by a power-law in the 2-10 keV energy range, with a relative narrow distribution
of slopes: $\langle \Gamma \rangle = 1.8 \pm 0.2$ \citep{Nandra1994,Steffen2006,Tozzi2006,young2009}.
Thus, the hard slope of the Cosmic X-ray Background (CXRB) spectrum 
(well described by a power-law with photon
index $\Gamma_{\rm CXRB} \simeq 1.4$ at $E < 10$ keV), and the prominent peak observed 
at about 30 keV are best accounted
for by assuming that the majority of active galactic nuclei are in fact obscured 
\citep{Setti1989,Comastri1995}.

In the traditional 'unification by orientation' schemes, 
the baffling diversity of AGN observational classes is explained on the basis of the line-of-sight
orientation with respect to the axis of rotational symmetry of the system \citep{Antonucci1985,Antonucci1993,Urry1995}.
In particular, obscured and un-obscured AGN are postulated to be intrinsically
the same objects, seen from different angles with respect to a dusty, large-scale, 
possibly clumpy, parsec-scale medium, which 
obscures the view of the inner engine \citep{Elitzur2008,Netzer2008}.
According to the simplest interpretations
of such unification schemes, there should not 
be any dependence of the obscured AGN fraction with intrinsic luminosity and/or redshift.

In recent years, new generations of synthesis models of the CXRB 
have been presented \citep{Gilli2007,Treister2009,Akylas2012}, 
following the publication of increasingly larger and deeper 
X-ray surveys, and reducing the uncertainties in the absorbing column density distribution. 
When combined with the observed X-ray luminosity functions, they now
provide an almost complete census of Compton thin AGN (i.e., those obscured by
columns $N_{\rm H} < \sigma_{\rm T}^{-1}\simeq 1.5 \times 10^{24}$ cm$^{-2}$, where $\sigma_{\rm T}$ 
is the Thomson cross section). This
class of objects dominates the counts in the lower X-ray energy band, where
almost the entire CXRB radiation has been resolved into individual sources \citep{Worsley2005}.

The results on the statistical properties of obscured AGN from these studies are 
at odds with the simple 'unification-by-orientation' scheme. 
In fact, evidence has been mounting over the years that
the fraction of absorbed AGN, defined in different and often
independent ways, appears to be lower at higher nuclear luminosities
\citep{Lawrence1982,Ueda2003,Steffen2003,Simpson2005,Hasinger2008,Brusa2010,Bongiorno2010,Burlon2011,Assef2013}. 
Such an evidence, however, is not uncontroversial. As recently summarized by \citet[][and references therein]{Lawrence2010},
the luminosity dependence of the obscured AGN fraction, so clearly detected, especially in X-ray selected samples,
is less significant in other AGN samples, such as those selected on the basis of their extended, low frequency
radio luminosity \citep{Willott2000} or in mid-IR colors \citep{Lacy2007}. The reasons for these discrepancies are
still unclear, with \citet{Mayo2013} arguing for a systematic bias in the X-ray selection due to 
an incorrect treatment of complex, partially-covered AGN.
In general, one would like to rely on large, complete samples selected on the basis
of a robust tracer of bolometric luminosity (independently on the the level of obscuration). 
Theoretically, hard X-ray and mid-IR selection should satisfy this basic requirement, with the former having the advantage 
of a much lower level of contamination by stellar processes \citep[see e.g.][for a discussion]{Merloni2013}.  

Evidence for a redshift evolution of the obscured AGN fraction is even
more controversial. Large samples of X-ray selected objects
have been used to corroborate claims of positive evolution of the fraction of
obscured AGN with increasing redshift \citep{Lafranca2005,Treister2006,Hasinger2008}, as well as 
counter-claims of no significant evolution \citep{Ueda2003,Tozzi2006,Gilli2007}. 
More focused investigation on specific AGN sub-samples,
such as $z>3$ X-ray selected QSOs \citep{Fiore2012,Vito2013}, 
rest-frame hard X-ray selected AGN \citep{Iwasawa2012}, or 
Compton Thick AGN candidates \citep{Brightman2012} 
in the CDFS have also suggested an increase of the incidence of
nuclear obscuration towards high redshift. Of critical importance 
is the ability of disentangling luminosity and redshift effects in (collections of)
flux-limited samples and the often complicated selection effects at high redshift, both in terms of 
source detection and identification/follow-up.

In a complementary approach to these ``demographic'' studies (in which the incidence of obscuration
and the covering fraction of the obscuring medium
is gauged statistically on the basis of large populations), SED-based investigations
look at the detailed spectral energy distribution of AGN, and at the IR-to-bolometric flux ratio in particular,
to infer the covering factor of
the obscuring medium in each individual source \citep{Maiolino2007,Treister2008,Sazonov2012,Roseboom2013,Lusso2013}. These studies
 also found general trends of decreasing covering factors with increasing nuclear (X-ray or bolometric) luminosity, and
little evidence of any redshift evolution \citep{Lusso2013}. Still, the results of these SED-based
investigations are not always in quantitative agreement with the demographic ones. 
This is probably due to the combined effects of the uncertain physical properties (optical depth, 
geometry and topology) of the obscuring medium \citep{Granato1994,Lusso2013}, 
as well as the unaccounted for biases in the observed {\it distribution} 
of covering factors for AGN of any given redshift and luminosity \citep{Roseboom2013}.

Both the apparent decrease of the incidence of obscured AGN  with intrinsic luminosity, as well as
its possible redshift evolution, might be (and have been widely) 
considered an indirect signature of AGN feedback, in that powerful sources are
able to clean up their immediate gaseous environments, responsible for the nuclear obscuration,
more efficiently \citep[see e.g.][]{Archibald2002,Hopkins2006,Menci2008}.  
In such AGN feedback models, unification-by-orientation scenarios are superseded by so-called
``evolutionary'' ones, as the incidence of nuclear obscuration changes dramatically with time 
during the SMBH active phases. To account for this, 
different physical models for the obscuring torus
have been proposed, all including some form of radiative coupling between the central
AGN and the obscuring medium \citep[e.g.][]{Lawrence1991,Maiolino2007,Nenkova2008}.

Irrespective of any specific model, it is clear that a detailed physical assessment of the interplay 
between AGN fuelling, star formation and
obscuration on the physical scales of the obscuring medium 
is crucial to our understanding of the mutual
influence of stellar and black hole mass growth in galactic nuclei \citep{Bianchi2012}.
Conceptually, we can identify three distinct spatial regions in the nucleus of a galaxy
on the basis of the physical properties of the AGN absorber. The outermost one is the gravitational
sphere of influence of the supermassive black hole (SMBH) 
itself, also called Bondi Radius 
$R_{\rm B} = 2GM_{\rm BH}/\sigma^2\simeq 10\, M_{\rm BH,8}\, \sigma_{\rm ,300}\; {\rm pc}$, where $M_{\rm BH,8}$ is the black hole mass in units of $10^{8}M_{\odot}$, and
$\sigma_{\rm ,300}$ can be either the velocity dispersion of stars for a purely collisionless nuclear environment, or
the sound speed of the gas just outside $R_{\rm B}$, measured in units of 300 km/s. 
To simplify, one can consider any absorbing
gas on scales larger than the SMBH sphere of influence to be ``galactic'', in the sense that
its properties are governed by star-formation and dynamical processes operating at the galactic scale. 
The fact that gas in the host galaxy can obscure AGN is not only predictable, but also clearly observed, 
either in individual objects (e.g. nucleus-obscuring dust lanes, \citealt{Matt2000}), or in larger samples
showing a lack of optically selected AGN in edge-on galaxies \citep{Maiolino1995,Lagos2011}.
Indeed, if evolutionary scenarios are to supersede the standard 
unification by orientation scheme and obscured AGN truly represent
a distinct phase in the evolution of a galaxy, then we expect
a relationship between the AGN obscuration distribution and the
larger scale physical properties of their host galaxies.

Within the gravitational sphere of influence of a SMBH, the most critical scale is the radius
within which dust sublimates under the effect of the AGN irradiation. A general treatment
of dust sublimation was presented in \citet{Barvainis1987,Fritz2006}, and subsequently applied to sophisticated
clumpy torus models \citep{Nenkova2008} or to interferometric observations of galactic nuclei in the near-IR
\citep{Kishimoto2007}. For typical dust composition, the dust sublimation radius is expected to scale as  
$R_{\rm d}\simeq 0.4\, (L_{\rm bol,45}/10^{45})^{1/2}(T_{\rm sub}/1500K)^{-2.6}\; {\rm pc}$ \citep{Nenkova2008}, as indeed
confirmed by interferometry observations of sizable samples of both obscured and un-obscured AGN in the nearby
Universe \citep{Tristram2009,Kishimoto2011}. 
Within this radius only atomic gas can survive, and reverberation mapping measurements do suggest that indeed
the Broad emission Line Region (BLR) is located immediately inside $R_{\rm d}$ \citep{Netzer1993,Kaspi2005}.
 
The parsec scale region between $R_{\rm d}$ and $R_{\rm B}$ is the traditional location of the obscuring torus
of the classical unified model. On the other hand, matter within $R_{\rm d}$ may be dust free, 
but could still cause substantial
obscuration of the inner tens of Schwarzschild radii of the accretion discs, where 
the bulk of the X-ray emission is produced \citep{Dai2010,Chen2012}. Indeed, a number of
X-ray observations of AGN have revealed in recent years the evidence for gas absorption within the
sublimation radius. Variable X-ray absorbers on short timescales are quite common 
\citep{Risaliti2002,Elvis2004,Risaliti2007, Maiolino2010}, and the variability timescales clearly suggests
that these absorbing structures lie within (or are part of) the BLR itself.

\begin{figure}
\includegraphics[width=0.48\textwidth,clip]{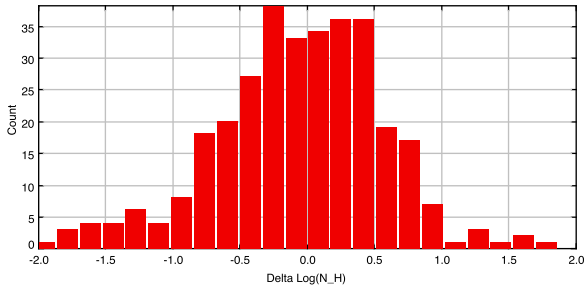}
\caption{Distribution of the difference between the logarithm of the best fit absorption column density $N_{\rm H}$ (in cm$^{-2}$)
between the HRz method and the full spectral analysis of \citet{Mainieri2011} for the 619 brightest sources in the sample.} 
\label{fig:deltanh}
\end{figure}

In light of all this bewildering complexity, general interpretations and global models for the inner structure
of a galaxy nucleus should be taken with extreme care, even very successful ones such as the 
'unification-by-orientation' scheme. Issues of sample selection, incompletenesses, or incomplete coverage of the 
nuclear Spectral Energy Distribution (SED) 
should all be accounted for in order to properly assess the overall relevance, for the 
AGN population at large, of any observational trend.

Here we present a study of the obscuration properties of a very large sample of 1310 AGN
selected in the XMM-COSMOS field \citep{Hasinger2007} on the basis of their X-ray luminosity.
In particular, we take advantage of the unprecedented level of identification and 
redshift completeness of the XMM-COSMOS \citep{Brusa2010,Salvato2009,Salvato2011} to
perform a {\it rest-frame 2-10 keV} X-ray flux selection, thus mitigating the $N_{\rm H}-z$ 
bias that plagues X-ray observed-frame flux limited samples \citep[see e.g.][]{Tozzi2006,Gilli2010b}.
The unique multi-wavelength coverage of the COSMOS
field \citep{Scoville2007} allows us to extract information on the host galaxies of the X-ray selected AGN.
We rely on the SED decomposition method of \citet{Merloni2010} and \citet{Bongiorno2012} to 
derive robust stellar mass estimates for all objects in the sample, and use the FIR
{\it Herschel-PEP} observations \citep{Lutz2011} of the field to infer star-formation rate indicators. 
Finally, the extensive spectroscopic followup granted through the zCOSMOS \citep{Lilly2007,Lilly2009} 
and Magellan/IMACS \citep{Trump2007,Trump2009} projects also plays a crucial role for the study we present here, 
together with archival data from the SDSS DR9 \citep{Ahn2012}.

The strength of our approach lies both in the sheer quantity and in the multi-wavelength
quality of the data in hand. Not only are we able to explore the relationships among 
AGN obscuration, nuclear luminosity and host galaxy stellar mass in complete bins 
of the luminosity-redshift parameter space with high statistical accuracy, but we can also
study the physical nature of the obscuration by comparing the classification of AGN into obscured and un-obscured
based on different indicators (X-ray spectral analysis, optical spectroscopy, optical-NIR SED).

The structure of the paper is as follows. In section~\ref{sec:sample} we describe the sample and the selection
process. In section~\ref{sec:frac_class} we explore the dependencies of the 
fraction of optically obscured AGN on luminosity, redshift and obscuration classification, while
in section~\ref{sec:obs_gal} we take a closer look at the relationship between obscured and un-obscured AGN and the 
stellar masses and star-formation rates of their hosts. Finally, in section~\ref{sec:x_vs_opt} we examine critically differences
and similarities in fraction of X-ray obscured AGN for the optical vs. the X-ray obscuration classification, before
drawing our conclusions in section~\ref{sec:conclusions}.

\begin{figure}
\includegraphics[width=0.48\textwidth,clip]{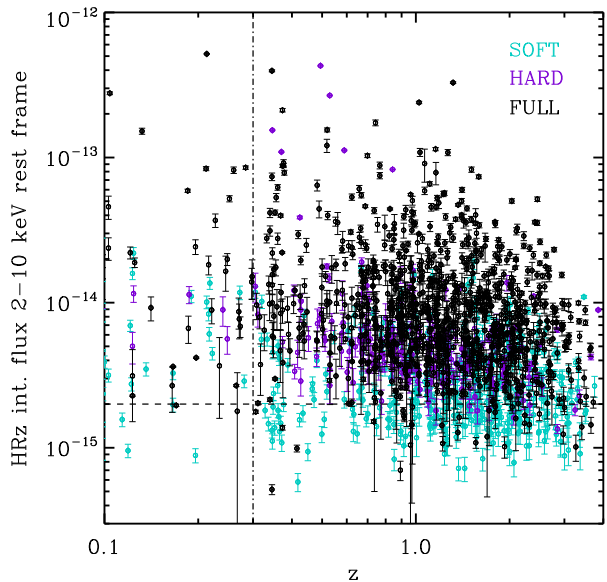}
\caption{Intrinsic (absorption corrected) rest-frame 2-10 keV fluxes
as a function of redshift for all the XMM-COSMOS AGN in the original
sample. Black, cyan and purple circles denote objects detected in the
full (0.5-2 and 2-10 bands), soft (0.5-2 keV) and hard (2-10 keV)
bands, respectively. The dashed and dot-dashed lines mark our
selection criteria for the analysis presented in this paper ($z>0.3$
and $f_{\rm 2-10  keV, rest-frame}=2 \times 10^{-15}$).} 
\label{fig:f210_rf}
\end{figure}

\section{Sample selection}
\label{sec:sample}

Our AGN were selected from the  XMM-COSMOS point-like
source catalog \citep{Hasinger2007,Cappelluti2009}, that includes $\sim$1800 X-ray
sources detected above a flux limit of $\sim 5 \times 10^{-16}$, $\sim 3
\times 10^{-15}$, and $\sim 7 \times 10^{-15}$ erg cm$^{-2}$ s$^{-1}$ 
in the 0.5--2 keV, 2--10 keV, and 5--10 keV bands, respectively, and 
has been presented in
\citet{Cappelluti2009}, to which we also refer for a detailed discussion of the
X-ray data analysis. Thanks to the rich multi-wavelength
data in the field, almost complete reliable counterpart identification
($>$98\%) was achieved by \citet{Brusa2010}; a combination of spectroscopic data (for more than
half of the sources) and good-quality photometric redshifts
\citep{Salvato2009,Salvato2011} ensures redshift completeness close to 100\%. 
  

\subsection{Rest-frame flux selection and intrinsic luminosities}
\label{sec:rest_frame_sel}

The main goal of this work is to study the incidence of 
obscuration in a large sample of X-ray selected AGN across a wide
range of luminosity, host properties and, crucially, redshift.
However, because of the wide energy band-pass of XMM,
 the X-ray selected sample described in the previous section
is a mixture of objects that have been detected above the flux
thresholds in the various energy bands. However, obscuration due to
neutral gas affects different bands at different redshifts in different
ways, causing a severe ``obscuration-redshift''
bias. As a result, flux-limited  samples
preferentially pick up more obscured objects at higher redshift
 \citep[see e.g.][]{Gilli2010b}.

Fortunately, the AGN sample at our disposal allows us to
circumvent (to a large extent) such a bias, as redshift is known for
all sources in the sample: a simple K-correction of the observed
spectrum can in principle guarantee a clean selection based on the
{\it rest-frame} flux in any given band. In the following, we call
this method for estimating rest-frame fluxes and luminosity on the
basis of the observed hardness ratios (HR) and redshifts (z) the 'HRz' method 
\citep{Ueda2003}.

We proceed with the HRz in the following way. 
For the 619 brightest sources (those with more than 200 pn counts in the full 
0.5-10 keV observed XMM band) we used the full spectral analysis of \citet{Mainieri2011}, and
obtain a value of the intrinsic column density $N_{\rm H}$ and 
rest-frame 2-10 keV flux (both observed and intrinsic).
For all other sources, we compute the
``observed'' spectral
slope ($\Gamma_{\rm obs}$) that best reproduces the ratio of X-ray
fluxes\footnote{In each band, the fluxes had been derived from the observed counts using fixed conversion factors, as
described in \citet{Cappelluti2009}.} in the soft
(0.5--2 keV) and hard (2--10 keV) X-ray bands. For objects detected only in one band, of course,
we can only place an upper (if only a 2-10 keV
detection is available) or a lower (if only a 0.5-2 keV
detection is available) limit on the observed spectral slope. In these cases we
assign each source a value of $\Gamma_{\rm obs}$ by picking
from the tails of the observed distribution of spectral slopes among
sources detected in both bands. 

The intrinsic (absorption corrected) luminosity for the AGN in the
sample was computed as in Brusa et al. (2010), i.e. 
on the basis of the the full spectral analysis \citep[see][]{Mainieri2011},
for objects with enough X-ray counts, while for the rest we use the HRz
method to estimate a value of the absorbing column density $N_{\rm H, HRz}$
in a statistical fashion, 
assuming each object has an intrinsic spectral slope which is normally distributed around a mean
of $\Gamma_{\rm int}=1.8 \pm 0.2$ (1$\sigma$ error)\footnote{All objects for which 
the inferred column density is smaller than an average galactic value of $3\times 10^{20}$ cm$^{-2}$ 
is assigned a value ${\rm Log} N_{\rm H}=20$}. The HRz method can be tested against full spectral analysis for the 
objects with the highest number of counts. The distribution of the difference in the logarithm of
the best fit column densities is plotted in Figure~\ref{fig:deltanh}. There are no apparent biases, although
the scatter is substantial, but only slightly larger than the statistical uncertainty on the spectrally
derived column densities, which has a median of about 0.3 dex.

\begin{figure}
\includegraphics[width=0.48\textwidth,clip]{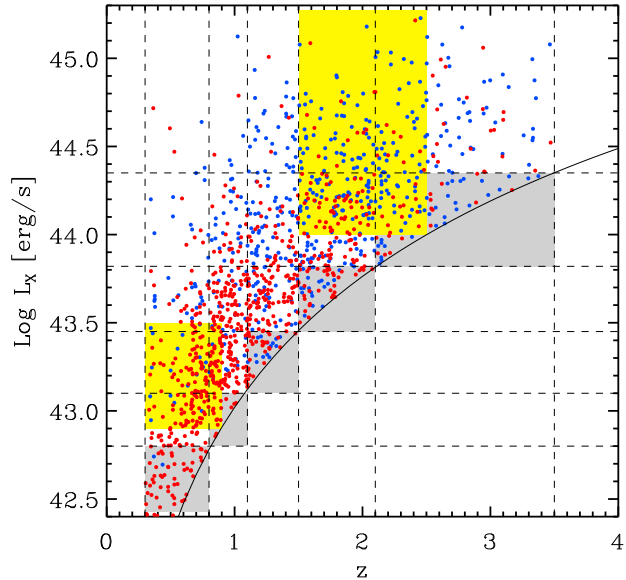}
\caption{Intrinsic (absorption corrected) 2-10 keV
  luminosity-redshift plane distribution of the sources studied. Blue
  circles represent optically classified type--1 AGN, red ones type--2 AGN. The vertical
  and horizontal dashed lines show the binning adopted for subsequent
  studies of complete samples. The grey shaded area are incomplete bins, while the yellow regions 
define the two luminosity-redshift interval used to investigate the optical/X-ray misclassified
AGN (see section~\ref{sec:stack} below).} 
\label{fig:lx_z}
\end{figure}

 The choice of a flux limit that defines our final
sample is somewhat arbitrary, but has to be chosen on the basis of the
best possible compromise between completeness and total number of
sources in the sample. Figure~\ref{fig:f210_rf} shows the rest-frame intrinsic
2-10 keV fluxes estimated with the HRz method as a function of redshift for the sources in the
original sample. We chose to strike this compromise at $f_{\rm 2-10
  keV, rest-frame}=2 \times 10^{-15}$ erg/s/cm$^2$, 
which is the intrinsic flux of the faintest sources detected in the hard band only 
(and thus most heavily obscured), further restricting our
analysis to sources at $z>0.3$, discarding only very few low-redshift
genuine AGN. We tested the robustness of such a choice by applying more conservative flux limits, up to
twice as bright a limit (i.e. up to $4 \times 10^{-15}$ erg/s/cm$^2$), but no qualitative change to the 
final results could be seen, apart from the obvious reduction of their statistical significance.

After excluding potential starburst galaxy contaminants (i.e. X-ray sources with
2-10 keV luminosity smaller than $2 \times 10^{42}$ erg/s) we are left
with a final sample of 1310 AGN selected on the basis of their
intrinsic (absorption corrected) rest-frame 2-10 keV flux at $z>0.3$.

\begin{figure*}
\includegraphics[width=0.45\textwidth,clip]{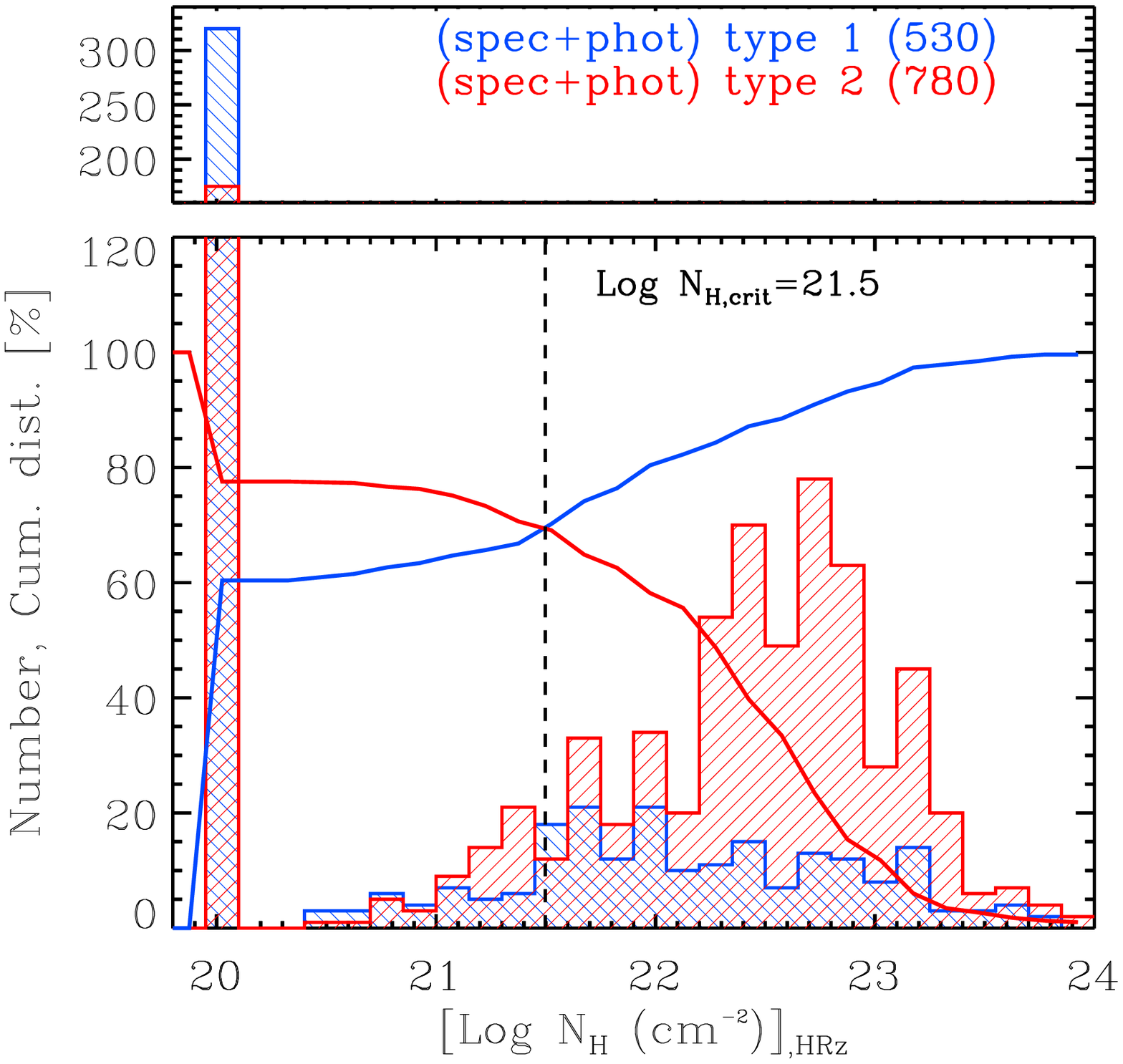} 
\includegraphics[width=0.45\textwidth,clip]{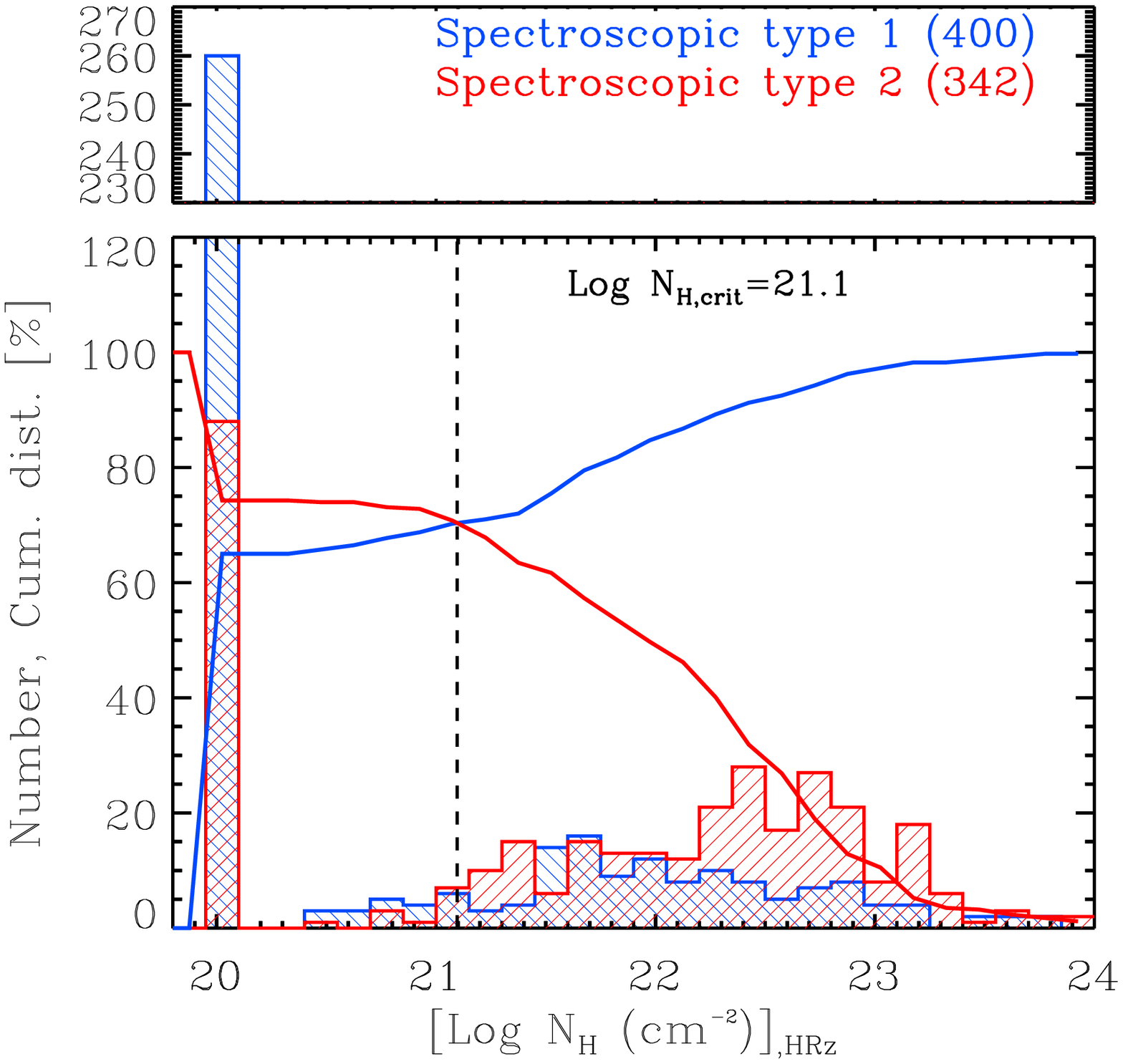} 
\caption{Distribution of the estimated column density (in logarithmic units) with the HRz method. 
  Blue shaded histograms are for optically classified type--1 AGN, red ones for type--2. 
  The {\it Left} panel is for the whole sample (spectroscopic and photometric; 1310 sources); the {\it Right} 
  panel for the spectroscopic redshift sample (742 sources in total). In each panel the
  blue solid lines are the cumulative distribution $C({\rm Log} N_{\rm H})$ (percentage) for the type--1
  AGN, and the red solid line is $1-C({\rm Log} N_{\rm H})$ for the type--2. 
 The vertical dashed lines indicates the values of ${\rm Log} N_{\rm H}$
 that best separates the two samples.} 
\label{fig:nh_dist}
\end{figure*}

We denote $L_{\rm X}$ the {\it intrinsic},
absorption-corrected 2-10 keV luminosity of the AGN. Because of the probabilistic way of assigning an intrinsic
value for the spectral index to each source, our final rest-frame flux limited sample is statistical in nature:
for the faint AGN detected in XMM-COSMOS, the individual values of$L_{\rm X}$, 
$N_{\rm H}$ and $\Gamma_{\rm int}$ are not precisely known,
but their ensemble average is accurate in a statistical sense. The sheer size of the sample then guarantees the robustness of
our results. Figure~\ref{fig:lx_z} shows the distribution of the sources in the
$L_{\rm X}-z$ plane. In the following sections we will describe the main results of our
analysis, performed in many instances by defining regions in the
luminosity-redshift plane in which our sample is complete; 
incomplete bins in the $L_{\rm X}-z$ plane
are shaded in Figure~\ref{fig:lx_z}.

One final note of caution is in place here. Such a simple approach for the absorption correction (i.e. the assumption that all
objects can be corrected for by taking into account a uniform, full-covering, absorber) might be
inadequate for more complex spectra, such as those produced by partial covering absorbers. These are
known to be present in many local AGN \citep{Winter2009}, albeit at luminosities typically lower than those we are
probing with our XMM-COSMOS sources. Such spectra are very hard to model properly  in survey data with
limited photon statistics \citep[but see][]{Brightman2012a}, but could dramatically affect our estimates
of the sources' intrinsic luminosity, for objects with Compton Thick absorbers covering a high fraction of the 
sky seen by the X-ray emitter. If this kind of objects represents a non-negligible fraction of 
the overall AGN population, the resulting effects on the statistics of obscured AGN fraction can be profound, as 
demonstrated by \citet{Mayo2013}. This is not predicted by current synthesis models of the CXRB, which all 
imply very small fractions of CT AGN at the flux limits of XMM-COSMOS \citep[see e.g.][]{Akylas2012}, 
but should obviously be checked with detailed spectral analysis of individual sources. Here, lacking the 
data quality sufficient for a proper treatment of
partially-covered AGN, we compare the 2-10 keV intrinsic luminosity we derived with the 
mid-IR luminosity (rest frame 12$\mu$m, measured from the overall optical/NIR SED fit, see 
Bongiorno et al. 2012).

Using a sample of local Seyfert galaxies (both Seyfert-1 and
Seyfert-2) at high spatial resolution, 
\citet{Gandhi2009} found that the uncontaminated nuclear mid-IR
continuum of AGN closely correlates with the intrinsic X-ray  [2--10] keV AGN
emitted powers over three orders of magnitude in luminosity, following the relation
$Log(L_{\rm MIR}/10^{43})=(0.19\pm0.05)+(1.11\pm0.07) Log(L_{\rm X}/10^{43})$. 
We show the $L_{\rm X}$-$L_{\rm MIR}$ for all the objects in our sample in 
Figure~\ref{fig:lir_lx_all}, color-coded by their absorption classification 
(see section~\ref{sec:sample_obs} below). Our objects cluster closely along the 
Gandhi et al. relation, with a noticeable scatter. More importantly, only a handful
of them have X-ray luminosities more than a factor of 10 smaller than what would be predicted 
on the basis of the 12$\mu$m luminosity. If our absorption corrections weren't accurate
(as in the case in which most of the objects were buried, Compton Thick AGN), 
we would have observed much higher $L_{12\mu{\rm m}}/L_{\rm X}$ ratios.
We conclude that, given the current limitations of the photon statistics, our simplified
spectral analysis is justified. We cannot exclude that some objects might be
affected by partial covering from a Compton Thick medium, but this should not
qualitatively change the overall results we present below.

\begin{figure}
\includegraphics[width=0.49\textwidth,clip]{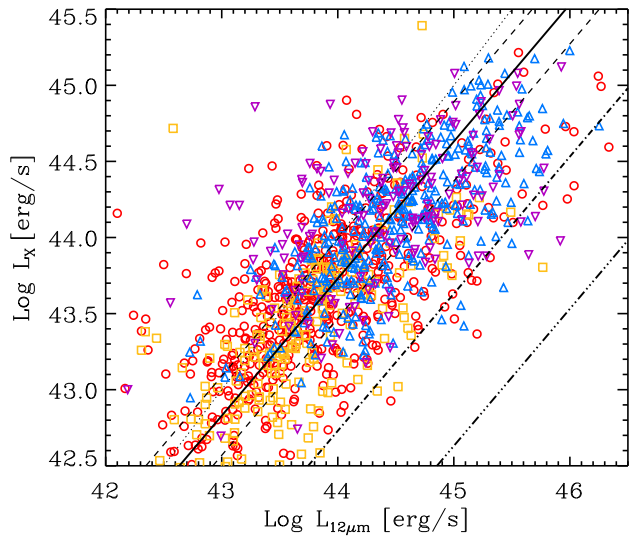}
\caption{Intrinsic (absorption corrected) 2-10 keV X-ray luminosity vs. rest-frame 12$\mu$m luminosity
(as derived from the optical/NIR SED fitting) for all AGN in the sample. Different colors and symbols represent 
objects with different optical and X-ray classification, as discussed in more detail in section~\ref{sec:x_vs_opt} below.
Red circles are for objects classified as obscured from both optical and X-ray spectra; blue upward triangles are 
un-obscured AGN from both optical and X-ray spectra; orange squares are optical type-2, but X-ray un-obscured, while
purple downwards triangles are optical type-1, but X-ray obscured objects. The solid line is the best fit relation from
Gandhi et al. (2009), with the dashed lines marking its 1-$\sigma$ scatter. The dot-dashed and triple-dot-dashed lines 
mark the locus where X-ray luminosity is a factor of 10 and 100 lower than predicted by the Gandhi relation, respectively.
The dotted line is the equality line, drawn to guide the eye only.} 
\label{fig:lir_lx_all}
\end{figure}

\subsection{Obscuration classification and host galaxy
  properties}
\label{sec:sample_obs}

The sources in the final sample can
be classified as obscured and un-obscured AGN
according to their optical/NIR properties, as discussed in
\citet{Brusa2010} and \citet{Bongiorno2012}. First of all, for all the
AGN with spectroscopic redshift information (742/1310 in our final sample), we classified as
optically un-obscured (type--1) those (400) which show broad optical emission lines (FWHM$>$
2000 km s$^{-1}$) and as optically obscured (type--2) all those (342) without broad
emission lines. For the remaining 568 sources, for which
only photometric redshifts
are available, the source classification in type--1 and type--2  
was performed on the basis of the template that best describes their
SED as derived from \citet{Salvato2011}; we have 130 ``photometric''
type--1 AGN and 438 ``photometric'' type--2. Photometric classification
relies on observables such as point-like morphology in the HST/ACS images \citep{Koekemoer2007}, blue rest-frame optical/UV colors and
significant variability to identify un-obscured AGN \citep{Salvato2009,Lusso2010} 
(for more details, see the flowchart in figure 8 of Salvato et al 2011).

As a first test of the different classification methods, 
we compare the column density estimated from the
X-ray broad-band spectra (HRz method) with the optical
classification in terms of type--1 and type--2 AGN described above. 
This is shown in Figure~\ref{fig:nh_dist} both for the
entire sample (left panel) and for the sub-sample of AGN with
spectroscopic redshift information (right panel). As expected, type--1 AGN (optically
un-obscured) have a column density distribution  which is peaked at values consistent with 
little or no X-ray
obscuration, but with a substantial tail at higher column densities. 
On the contrary, most type--2
(optically obscured) AGN appear to have column densities ${\rm Log} N_{\rm H}>21$, but
 again with a non-negligible
fraction of sources (both spectroscopically and photometrically
classified) with X-ray spectra (or hardness ratios) consistent with no
obscuration.


\begin{figure*}
\includegraphics[width=0.49\textwidth,clip]{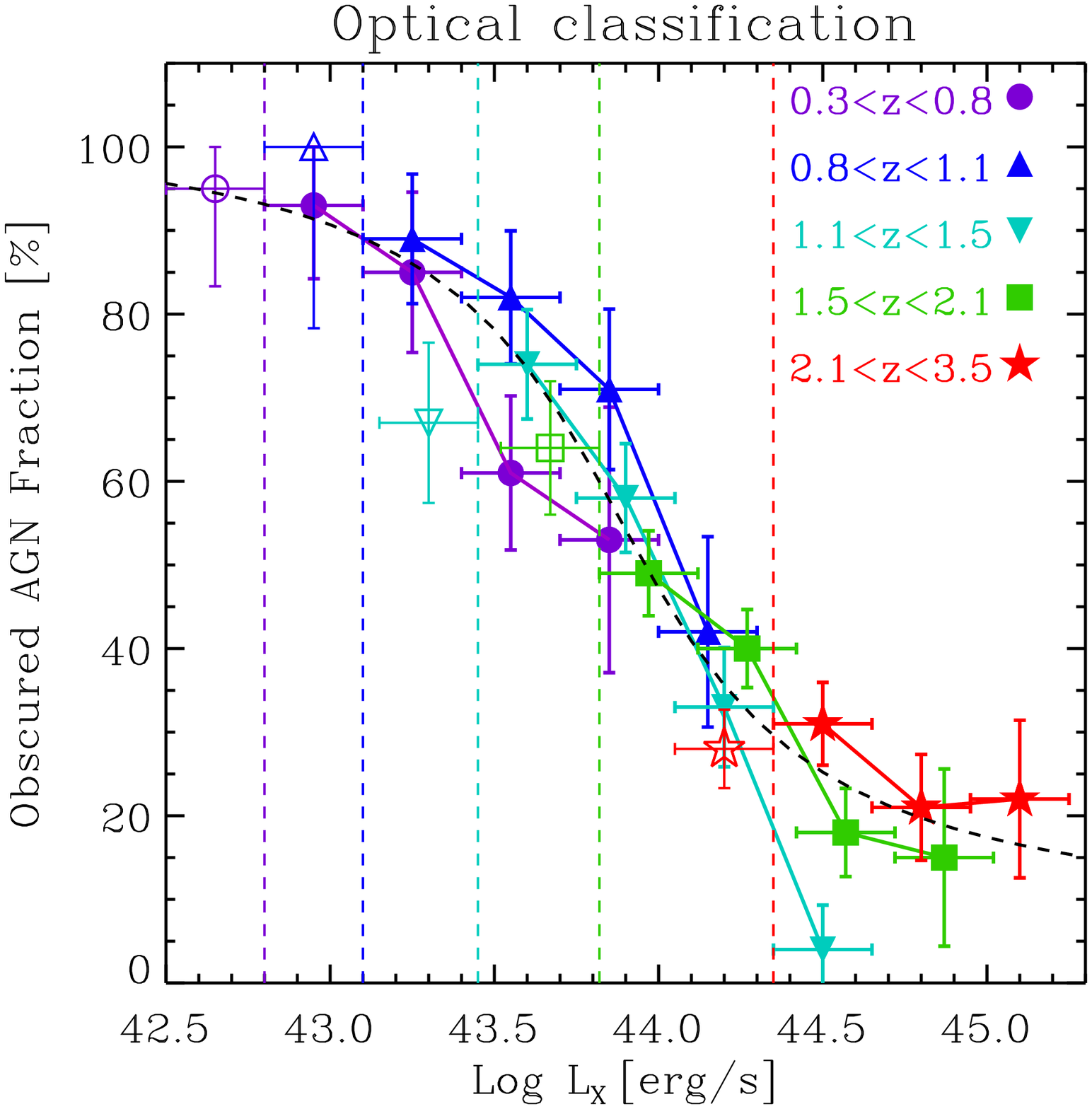}
\includegraphics[width=0.49\textwidth,clip]{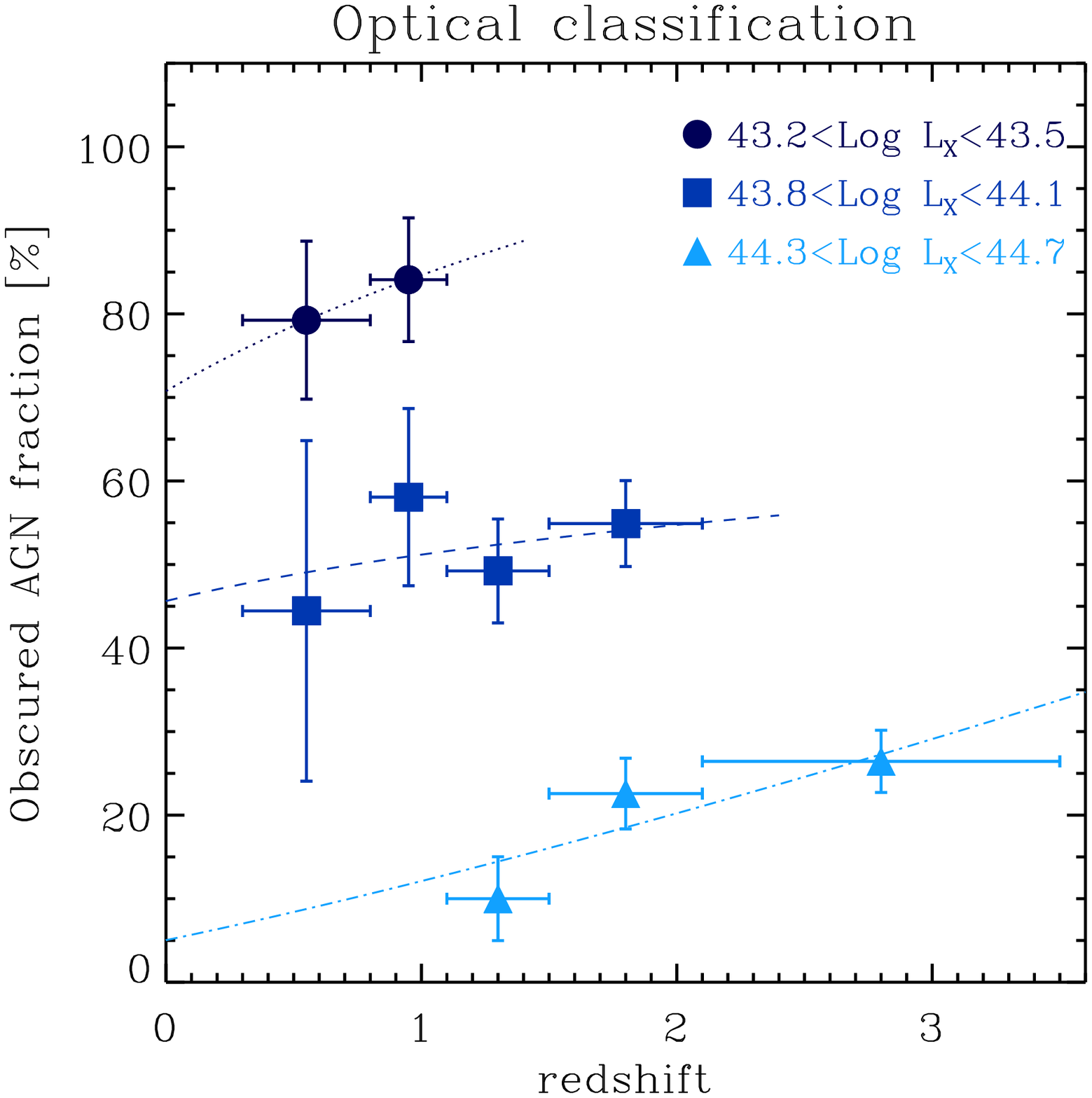}
\caption{{\it Left Panel}: The fraction of optically obscured AGN is plotted
  versus the X-ray luminosity for different redshift bins (purple
  circles: $0.3 \le z<0.8$; blue upwards triangles: $0.8 \le z<1.1$;
  cyan downwards triangles: $1.1 \le z<1.5$; green squares: $1.5 \le
  z<2.1$ and red stars: $2.1 \le z<3.5$). The vertical dashed lines
  mark the luminosities above which the samples are complete in each
  redshift bin (of corresponding color). Empty symbols are from
  incomplete bins. The
  dashed line is the best fit
  to the entire data set across the whole redshift range (see text for details).  
{\it Right Panel}: Redshift evolution of the fraction of Obscured AGN in different luminosity bins (only those for which we are 
complete have been shown). The dotted, dashed, and dot-dashed lines show the best fit evolution in the three luminosity interval, respectively,
according to Eq.(\ref{eq:z_evol}).
The best fit values for the parameter of each curve are reported in Table~\ref{tab:zev}.} 
\label{fig:abs_frac_lx_allz}
\end{figure*}

We can quantify this mismatch between the two
classification methods (optical- and X-ray-based) by comparing the
cumulative $N_{\rm H}$ distribution of the type--1 AGN (i.e. the cumulative number
of objects with column density {\it smaller} than a given value, blue curves in Fig.~\ref{fig:nh_dist}) 
with the opposite of
the cumulative $N_{\rm H}$ distribution of type--2 AGN (i.e. the cumulative number
of objects with column density {\it larger} than a given value, red curves in
Fig.~\ref{fig:nh_dist}). Their intersection tells us what value of
$N_{\rm H}$ best separates the two samples (measured at ${\rm Log} N_{\rm H}=21.5$ and ${\rm Log} N_{\rm H}=21.1$ 
for the full and the spectroscopic samples, respectively), while the level at which
the two curves intersect can be used to define a level of
``coherence'' of the two classifications, with unity being perfect
consistency. As we can see from both panels, typically the two
classifications agree to within about 70\%, i.e. there are about 30\%
of optically classified type--1 (type--2) AGN with (without) evidence of
absorption in the X-ray spectra. This is consistent with the similar analysis
done by \citet{Trouille2009} on three separate {\it Chandra} survey fields. 
In the remaining of this work, we will adopt a value of 
$N_{\rm H,crit}=10^{21.5}$ cm$^{-2}$ to distinguish between X-ray obscured and un-obscured AGN. Similar values 
for the discriminating column density had been
previously argued in studies of the optical properties of smaller hard X-rays selected samples
 \citep{DellaCeca2008,Caccianiga2008}.

It is worth noticing here that the values of the neutral hydrogen equivalent column density that best separate the
obscured and un-obscured samples, providing the maximal overlap between the X-ray- and optical-based classification
are smaller than the traditionally adopted value of $10^{22}$ cm$^{-2}$. Assuming a standard dust-to-gas ratio \citep{Predehl1995}
and Galactic extinction law, the critical column densities correspond to optical extinction values of $E(B-V)\simeq 0.57$ 
for the entire sample and $E(B-V)\simeq 0.23$ for the spectroscopic sample. These values of extinction are consistent with 
the findings of \citet{Lusso2010}, who have studied in detail the SED of spectroscopic type--1 AGN in COSMOS, concluding that 
only a negligible minority appears to suffer from reddening in excess of $E(B-V)=0.2$.

Finally, we derive the host galaxy properties of our AGN using the results of the analysis presented in
\citet{Bongiorno2012}, where we introduced a procedure to fit the multi-wavelength
SED of the sources by disentangling the
nuclear and stellar contributions.  In particular, as discussed in \citet{Bongiorno2012}, at rest-frame wavelengths
of about 1$\mu$m, three quarters of type--1 AGN have AGN fractions
smaller than about 50\%, implying that, even for un-obscured objects,
our method allows a robust determination of the host total stellar
masses, with typical uncertainty (statistical plus systematic) 
of the order of 0.4 dex (see also Merloni et al. 2010). Thanks to that
approach, stellar masses for the host galaxies of all the AGN
in our final sample could be obtained.

\section{The fraction of  obscured AGN: luminosity, redshift and classification dependencies}
\label{sec:frac_class}
First of all, we examined the luminosity and redshift dependence of the fraction of AGN
optically classified as obscured. The left hand panel of 
Figure~\ref{fig:abs_frac_lx_allz} shows such a fraction
as a function of intrinsic X-ray luminosity in five different redshift bins (see also Table~\ref{tab:abs_frac_opt}). 
The decrease of 
the obscured AGN fraction with luminosity is strong, and confirms previous studies on the XMM-COSMOS 
AGN \citep{Brusa2010}. 
The dashed line, instead, shows the best fit relations to the
optically obscured AGN fraction obtained combining all redshift bins. We assumed the relation to be of the form:
\begin{equation}
\label{eq:fit}
F_{\rm obs}=A+\frac{1}{\pi}atan\left(\frac{l_0-l_x}{\sigma_x}\right)
\end{equation}
where $l_x= {\rm Log} L_{\rm X}$. The best-fit parameters we obtained are: $A=0.56$; $l_0=43.89$ and 
$\sigma_x=0.46$.

As mentioned before, the strong trend we observe could, at least partially, be due to an incorrect
estimation of the intrinsic X-ray luminosity, which can induce an apparent trend of the obscured
AGN fraction with luminosity, even in the absence of a real one. \citet{Mayo2013} have shown that, in order
for such an effect to be dominant, a substantial fraction ($\sim$50-60\%) of all objects under study should be mis-modelled
partially-covered AGN, with Compton Thick absorbers covering more than 95\% of the X-ray source. This is difficult to 
rule out completely, but we note here that both the comparison with the estimated MIR luminosity, and the 
average X-ray spectral shape of the objects in our sample (see Figures~\ref{fig:lir_lx_all} and \ref{fig:stack_x}) 
suggest that the fraction of Compton Thick AGN in the sample is not as large.

\begin{table}
\caption{Fraction of optically (from figure~\ref{fig:abs_frac_lx_allz}) and X-ray (from figure~\ref{fig:abs_frac_lx_allz_hrz}) obscured AGN for various redshift and X-ray luminosity bins. Incomplete bins are marked with an asterisk.}
\label{tab:abs_frac_opt}
\begin{tabular}{ccc}
\hline
\multicolumn{3}{c}{$0.3 \le z < 0.8$}\\
\hline
\hline
${\rm Log} L_{\rm X}$ [erg/s] & Optical Obsc. Frac. [\%]  & X-ray Obsc. Fract. [\%]\\
\hline
42.5 -- 42.8 & $^*$95$^{+5}_{-12}$ & $^*$58$\pm 8$\\
42.8 -- 43.1 & 93$^{+7}_{-9}$ & 59$\pm 6$\\
43.1 -- 43.4 & 85$\pm 10$ & 54$\pm 7$\\
43.4 -- 43.7 & 61$\pm 9$ & 45$\pm 8$ \\
43.7 -- 44.0 & 53$\pm 16$ & 46$\pm 15$ \\
\hline
\hline
\multicolumn{3}{c}{$0.8 \le z < 1.1$}\\
\hline
\hline
${\rm Log} L_{\rm X}$ [erg/s] & Optical Obsc. Fract. [\%]  & X-ray Obsc. Fract. [\%]\\
\hline
42.8 -- 43.1 & $^*$100$^{+0}_{-22}$ & $^*$35$\pm 12$\\
43.1 -- 43.4 & 89$\pm 8$  & 68$\pm 6$\\
43.4 -- 43.7 & 82$\pm 8$ & 68$\pm 7$ \\
43.7 -- 44.0 & 71$\pm 10$  & 60$\pm 9$\\
44.0 -- 44.3 & 42$\pm 11$ & 47$\pm 12$\\
\hline
\hline
\multicolumn{3}{c}{$1.1 \le z < 1.5$}\\
\hline
\hline
${\rm Log} L_{\rm X}$ [erg/s] & Optical Obsc. Fract. [\%]  & X-ray Obsc. Fract. [\%]\\
\hline
43.15 -- 43.45 & $^*$67$\pm 10$ & $^*$46$\pm 8$ \\
43.45 -- 43.75 & 74$\pm 7$  & 68$\pm 6$\\
43.75 -- 44.05 & 58$\pm 7$  & 59$\pm 7$ \\
44.05 -- 44.35 & 33$\pm 7$ & 44$\pm 8$\\
44.35 -- 44.65 & 4$^{+5}_{-4}$ & 22$\pm 8$ \\
\hline
\hline
\multicolumn{3}{c}{$1.5 \le z < 2.1$}\\
\hline
\hline
${\rm Log} L_{\rm X}$ [erg/s] & Optical Obsc. Fract. [\%]  & X-ray Obsc. Fract. [\%]\\
\hline
43.52 -- 43.82 & $^*$64$\pm 8$ &  $^*$46$\pm 7$\\
43.82 -- 44.12 & 49$\pm 5$  & 56$\pm 6$ \\
44.12 -- 44.42 & 40$\pm 5$  & 55$\pm 6$\\
44.42 -- 44.72 & 18$\pm 5$ & 36$\pm 7$\\
44.72 -- 45.02 & 15$\pm 11$ & 53$\pm 17$\\
\hline
\hline
\multicolumn{3}{c}{$2.1 \le z < 3.5$}\\
\hline
\hline
${\rm Log} L_{\rm X}$ [erg/s] & Optical Obsc. Fract. [\%]  & X-ray Obsc. Fract. [\%]\\
\hline
44.05 -- 44.35 & $^*$28$\pm 5$ & $^*$49$\pm 6$\\
44.35 -- 44.65 & 31$\pm 5$ & 57$\pm 7$\\
44.65 -- 44.95 & 21$\pm 6$ & 53$\pm 10$\\
44.95 -- 45.25 & 22$\pm 9$ & 50$\pm 14$\\
\hline
\end{tabular}
\end{table}

\begin{figure*}
\includegraphics[width=0.49\textwidth,clip]{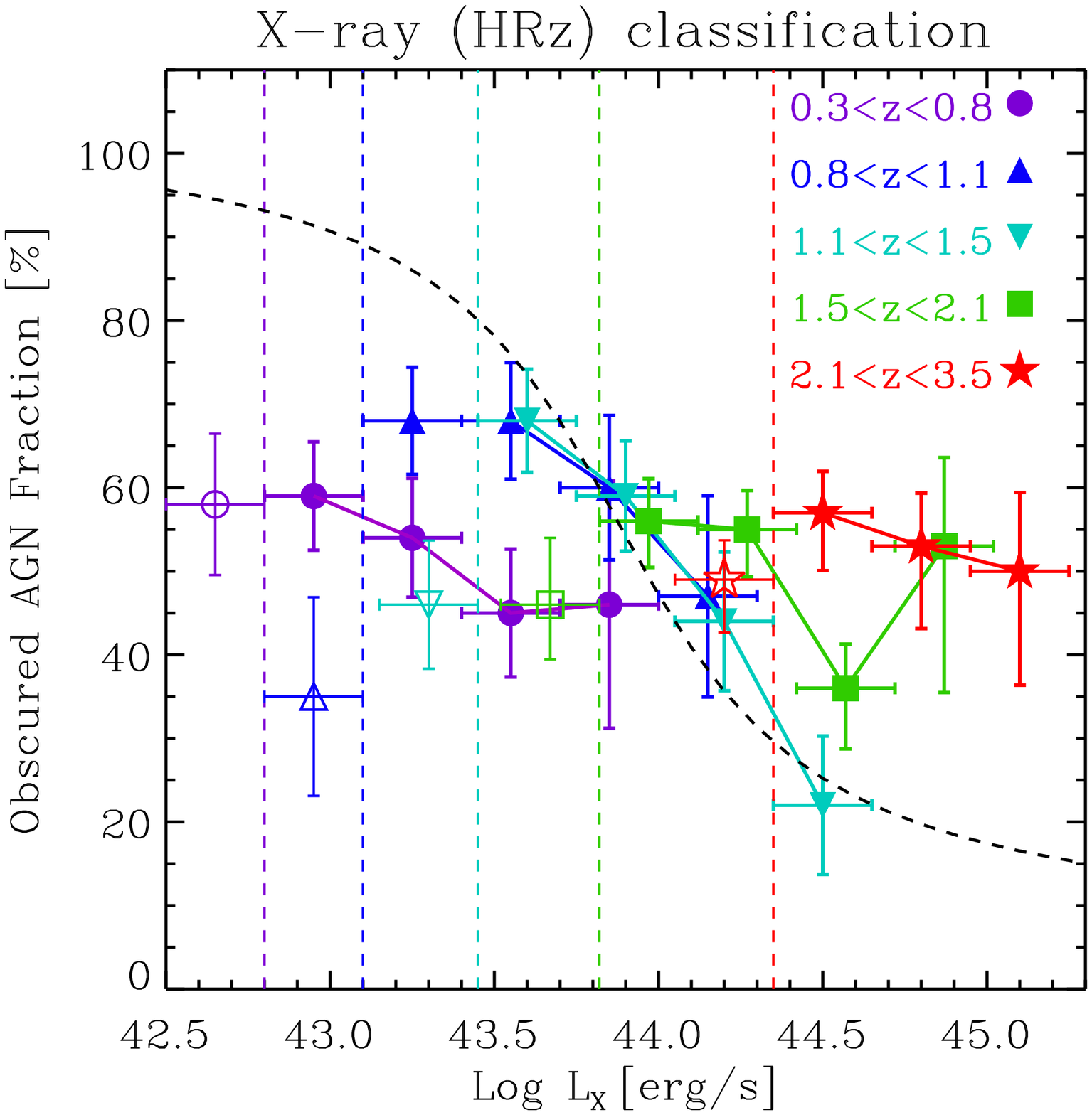}
\includegraphics[width=0.49\textwidth,clip]{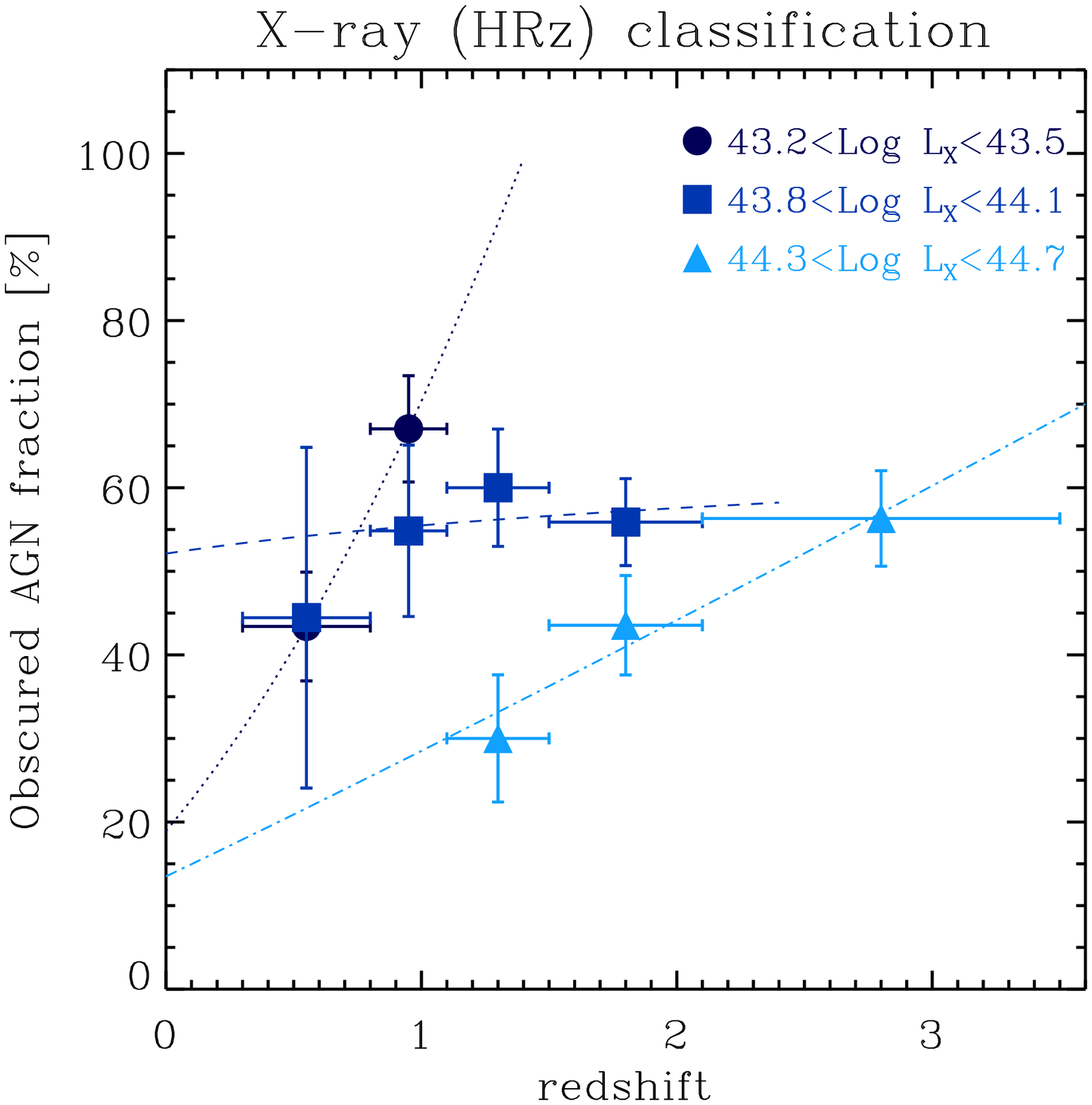}
\caption{The fraction of X-ray obscured AGN is plotted
  versus the X-ray luminosity for different redshift bins (purple
  circles: $0.3 \le z<0.8$; blue upwards triangles: $0.8 \le z<1.1$;
  cyan downwards triangles: $1.1 \le z<1.5$; green squares: $1.5 \le
  z<2.1$ and red stars: $2.1 \le z<3.5$). The vertical dashed lines
  mark the luminosities above which the samples are complete in each
  redshift bin (of corresponding color). Empty symbols are from
  incomplete bins. The dashed line is here plotted as a reference, and represent the best fit to the absorbed AGN fraction vs. luminosity
relation for optically obscured AGN, from Fig.~\ref{fig:abs_frac_lx_allz} and Eq.(\ref{eq:fit}).
{\it Right Panel}: Redshift evolution of the fraction of Obscured AGN in different luminosity bins (only those for which we are 
complete have been shown). The dotted, dashed, and dot-dashed lines show the best fit evolution in the three luminosity interval, 
respectively,
according to Eq.(\ref{eq:z_evol}).
The best fit values for the parameter of each curve are reported in Table~\ref{tab:zev}.} 
\label{fig:abs_frac_lx_allz_hrz}
\end{figure*}

In section~\ref{sec:sample_obs} we have shown how the classification based on the optical properties 
(spectroscopic and/or
photometric) of the AGN and that based on the X-ray spectral analysis (mostly through the HRz method) agree in no more
than 70\% of the sample. Because of this, it is mandatory to explore and analyse the differences in the 
statistical properties of the sample when using an alternative (X-ray based) classification scheme.
The left panel of Figure~\ref{fig:abs_frac_lx_allz_hrz} shows the fraction of (X-ray classified) obscured AGN as a function
of intrinsic 2-10 keV X-ray luminosity ($L_{\rm X}$) for different redshift intervals. Not in all redshift intervals
there is a clear decrease of the fraction of obscured AGN with increasing luminosity; in particular, in the highest redshift
bin the obscured AGN fraction remains almost constant, but the range of intrinsic luminosities sampled is too limited
to draw firmer conclusions.

We now turn to a discussion of any possible evolutionary trend of the obscured AGN fractions as a function of redshift.

\subsection{Redshift evolution}
We plot in the right panels of 
Figures~\ref{fig:abs_frac_lx_allz} and \ref{fig:abs_frac_lx_allz_hrz} the fraction of obscured AGN as a function of
redshift, for three separate luminosity intervals and for the optical and X-ray classifications, respectively. 
Only bins where the sample is complete are shown.
For optically classified AGN, we do not see any clear redshift evolution, apart from the highest luminosity objects 
(i.e. genuine QSOs in the XMM-COSMOS sample, with $L_{\rm X}$ between 10$^{44.3}$ and 10$^{44.7}$ erg/s). To better quantify this,
we have fitted separately the evolution of the obscured fraction in the three luminosity bins with the function:
\begin{equation}
F_{\rm obs}=B \times (1+z)^{\delta}.
\label{eq:z_evol}
\end{equation} 

The best fit parameters are shown in Table~\ref{tab:zev}, while the best fit relations are shown as thin lines in
the right panels of Figures~\ref{fig:abs_frac_lx_allz} and \ref{fig:abs_frac_lx_allz_hrz}. For the optical classification, as anticipated, 
we measure a significant evolution ($\delta_{\rm OPT} > 0$) 
only for the most luminous objects, with $\delta_{\rm OPT}=1.27\pm0.62$. Previously, \citet{Treister2006a} had claimed a significant increase in the fraction
 of optically classified obscured AGN with redshift, on the basis of the analysis of a large sample extracted from a combination
of seven different X-ray surveys of various areas and depth. 
Their sample indeed covers a wider luminosity and redshift range, so is in principle better
suited to disentangle their degeneracies, 
but, on the other hand, requires corrections to be made for compensating for the redshift incompleteness 
of the original X-ray selected sample, which could introduce biases, if the spectroscopic redshift completeness correlates with obscuration.
Our nearly complete sample (within the volumes probed by the COSMOS field 2 deg$^2$ area), is at least minimally affected by such bias.
 
For the X-ray classification, we observe a significant evolution with redshift 
both  at the lowest and highest luminosities, where the fraction of X-ray obscured AGN increases with $z$, consistent with previous findings
by numerous authors
 \citep[][]{Lafranca2005,Treister2006a,Hasinger2008,Fiore2012,Vito2013}. 
A more robust assessment of the redshift evolution 
of the obscured AGN fraction, however, would require a more extensive coverage of the $L-z$ plane than that afforded by our 
flux-limited XMM-COSMOS sample. 

\begin{table}
\caption{Best fit parameter (from equation \ref{eq:z_evol}) for the redshift evolution of the obscured AGN in three different luminosity bins. The parameters for the optically classified AGN are marked by 'OPT', while those for the X-ray classified AGN by 'HRz'. In boldface,
we highlight the sub-sample where significant redshift evolution is seen.}
\label{tab:zev}
\begin{tabular}{cccc}
\hline
&\multicolumn{3}{c}{${\rm Log}L_{\rm X}\;\; {\rm range}$}\\
${\rm parameter}$ & $[43.2,43.5]$ & $[43.8,44.1]$ & $[44.3,44.7]$ \\
\hline
$B_{\rm OPT}$ & 0.71$\pm$0.27 & 0.46$\pm$0.20 & {\bf 0.05$\pm$0.04} \\
$\delta_{\rm OPT}$ & 0.26$\pm$0.65 & 0.17$\pm$0.46 & {\bf 1.27$\pm$0.62} \\
\hline
$B_{\rm HRz}$ & {\bf 0.18$\pm$0.09} & 0.52$\pm$0.21 & {\bf 0.13$\pm$0.07} \\
$\delta_{\rm HRz}$ & {\bf 1.89$\pm$0.77} & 0.09$\pm$0.43 & {\bf 1.08$\pm$0.42} \\
\hline
\hline
\end{tabular}
\end{table}

\section{Obscured AGN and galaxy-wide properties}
\label{sec:obs_gal}

\begin{figure}
\includegraphics[width=0.48\textwidth,clip]{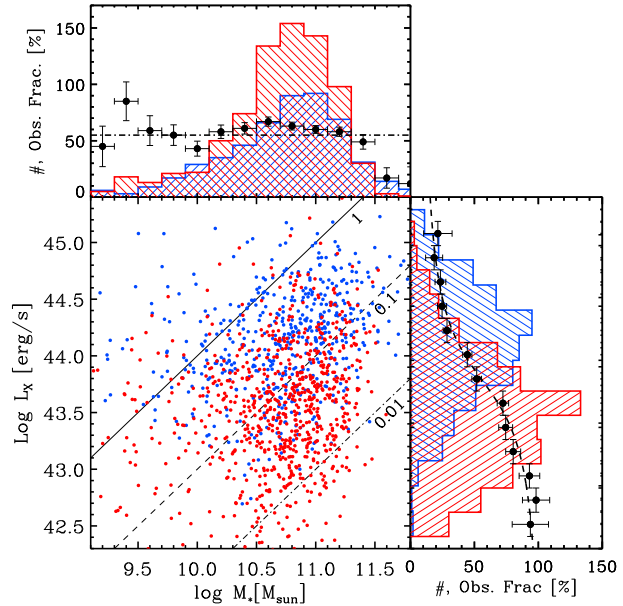}
\caption{The main panel shows the distribution of the 1310 AGN in the sample in the ${\rm Log}\, M_*$-${\rm Log}\,L_{\rm X}$
plane. Blue circles are optically classified type--1 objects, red type--2 objects. The dot-dashed, dashed, and 
solid lines mark the approximate locus
of 1\%, 10\% and 100\% of the Eddington limit, estimated by assuming constant 2-10 keV bolometric correction, 
$\kappa_{\rm X}$ and a universal $M_{\rm BH}/M_*$ ratio, such that $\kappa_{\rm X}M_*/M_{\rm BH}=10^{3.9}$. 
The right hand panel shows the luminosity distributions
for the two classes, with over-plotted the obscured AGN fraction (black circles with error bars) and the best 
fit relation of eq.(\ref{eq:fit}). The top panel shows the distribution of the host galaxies' stellar masses for the 
two classes of AGN, with over-plotted the obscured AGN fraction (black circles with error bars) calculated for different mass bins; 
in each of these fixed stellar mass bins, the 
obscured AGN fraction is calculated mixing AGN of different luminosities, and the result is an almost mass-independent average
obscured fraction of about 55\% (dot-dashed horizontal line).}
\label{fig:lx_m}
\end{figure}

\subsection{Relationship between AGN obscuration and host galaxies stellar masses}
An important test that is made possible by the quality of the data of the COSMOS AGN sample is
to study whether the host galaxy stellar mass ($M_*$) plays any role in determining the statistics of
(optically) obscured AGN in our sample. Figure~\ref{fig:lx_m} shows the distribution of the 1310 AGN in the sample 
in the ${\rm Log}\,L_{\rm X}$-${\rm Log}\, M_*$ 
plane. Most of the AGN hosts in our sample have stellar masses within one decade 
(between $10^{10.3}$ and $10^{11.3}$ solar masses),
but span a large range of ``specific accretion rates''
\citep{Aird2012,Bongiorno2012}, i.e. the observed ratio of nuclear X-ray luminosity to host galaxy
stellar mass. Modulo the AGN bolometric correction, the standard scaling relation between $M_*$ and 
the SMBH mass \citep{Haring2004} 
would imply that the specific accretion rate could be used as a rough proxy of the Eddington ratio, and
Figure~\ref{fig:lx_m} shows that most AGN in our sample accrete at Eddington ratios of the order of 10\%, but with a 
very wide distribution.  
The distribution of the host galaxies' stellar masses for the 
two classes of AGN, with over-plotted the obscured AGN fraction (black circles with error bars) calculated for different mass bins is shown 
in the top panel of Figure~\ref{fig:lx_m}. In each of these bins, the 
obscured AGN fraction is calculated mixing AGN of different luminosities, and the result is an almost mass-independent average
obscured fraction of about 55\%. It is interesting to note that a number of authors have
previously noticed that the overall fraction of obscured AGN is a universal number, indeed close to this value \citep{Lawrence2010}.  
The difference with our findings could in part be due to the effect of applying very uncertain (and intrinsically scattered) 
bolometric corrections to a compilation of data-sets from different catalogs. The effect of this would be to wash out the true 
luminosity dependence and average the obscured AGN fraction to a value close to the sample mean of $\sim$55\%. Alternatively, the 
problem could lie in the derivation of the intrinsic luminosity from the observed X-ray one, which could introduce a spurious 
luminosity dependence, as discussed in sections~\ref{sec:rest_frame_sel} and \ref{sec:frac_class} above.

\begin{figure*}
\includegraphics[width=0.45\textwidth,clip]{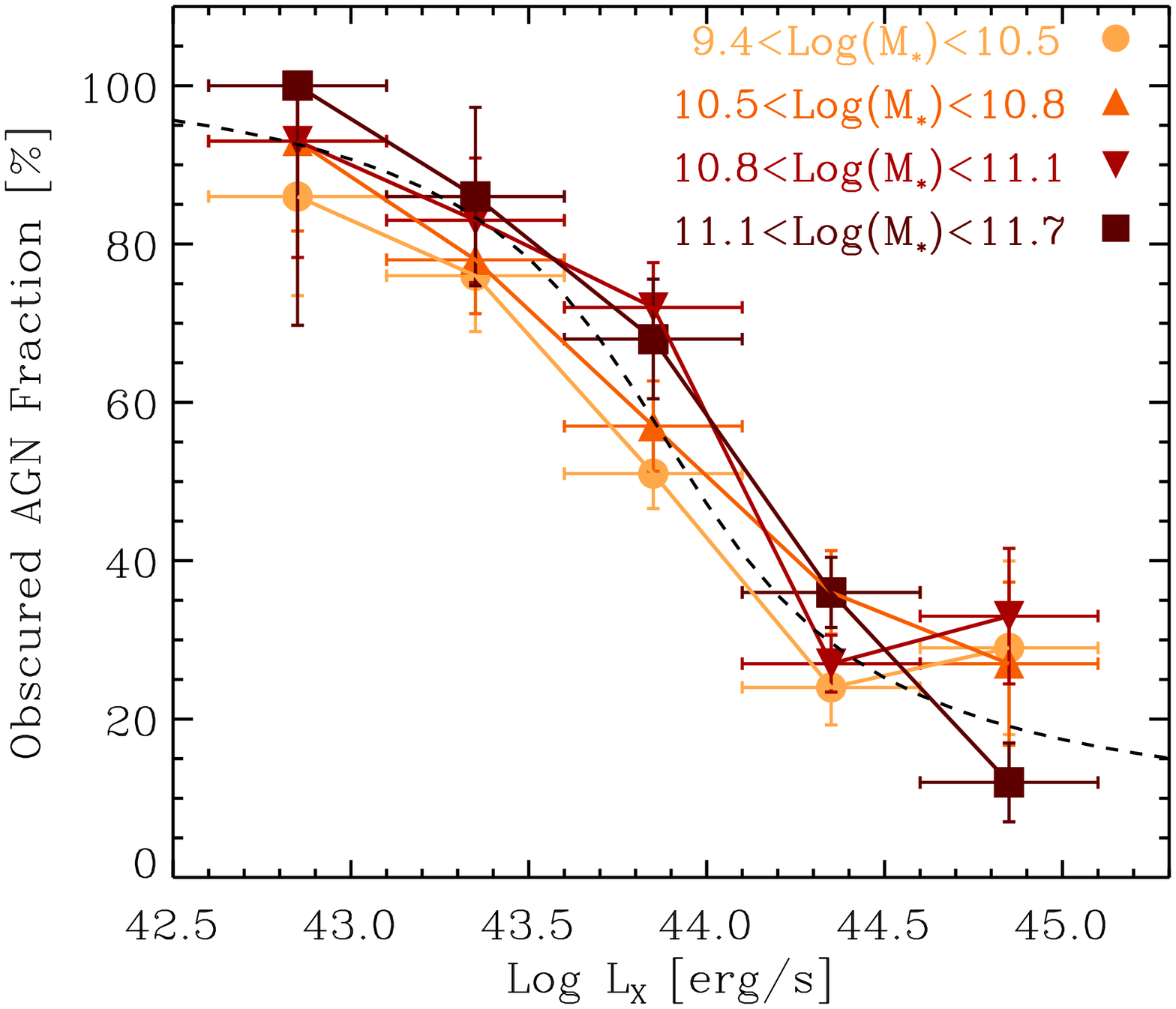}
\includegraphics[width=0.45\textwidth,clip]{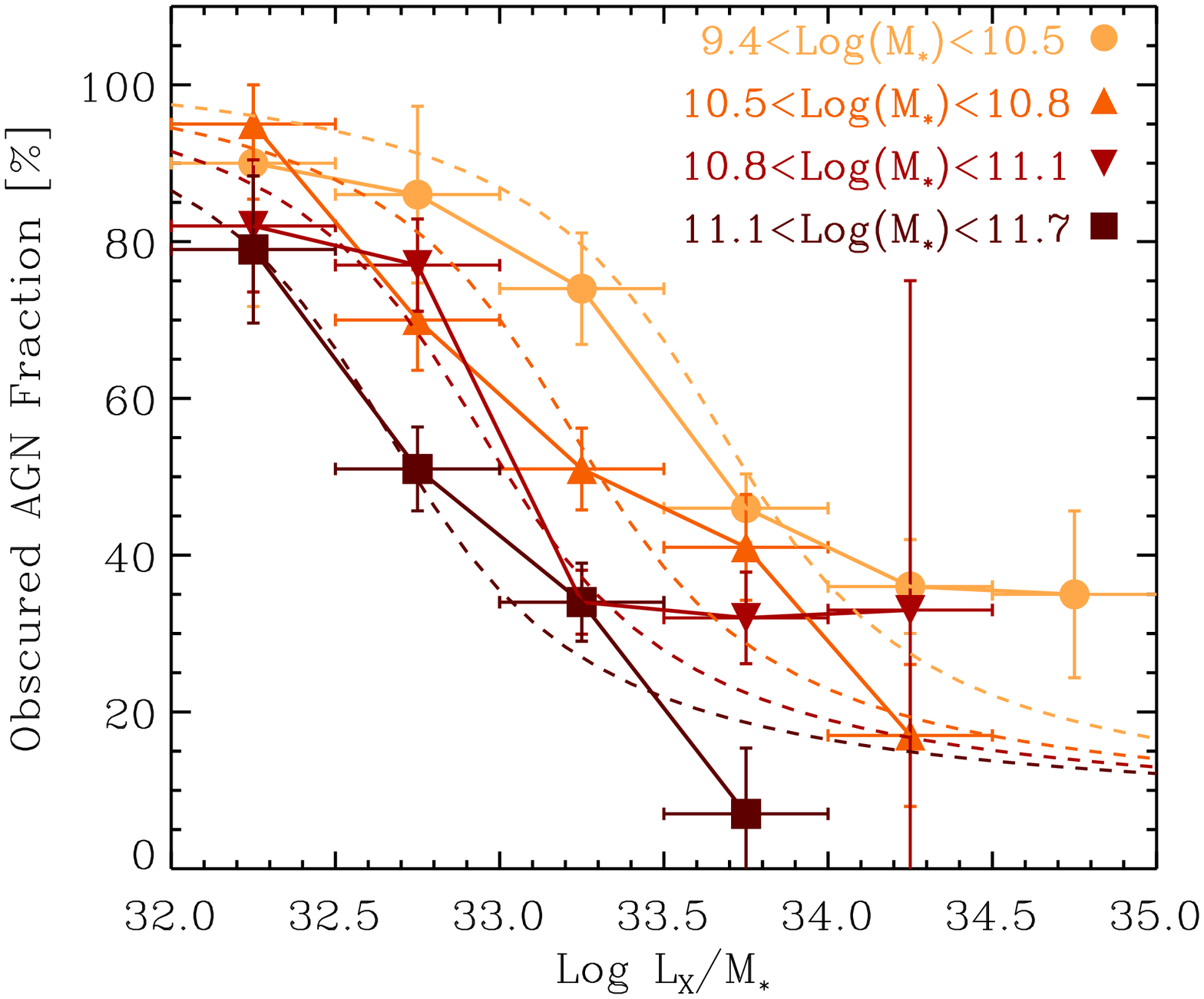}
\caption{{\it Left:} The fraction of obscured AGN is plotted
  versus the X-ray luminosity for different host galaxy
  stellar mass bins. The dashed line is the best fit 
  to the entire data set shown in Fig.~\ref{fig:abs_frac_lx_allz}. {\it Right:} The fraction of 
obscured AGN is plotted 
  versus the specific accretion rate ($L_{\rm X}/M_{*}$), for different host galaxy
  stellar mass bins. The
  dashed lines are the best fit of the 
  to the entire data set shown in the left panel of Fig.~\ref{fig:abs_frac_lx_allz}, each
shifted horizontally by the corresponding factor of ${\rm Log} M_{*}$.}. 
\label{fig:abs_frac_lx_mass}
\end{figure*}

Despite its size, our sample is still too small to study
simultaneously the obscured AGN fraction as a function of luminosity, redshift and stellar mass. However, 
the results of Fig.~\ref{fig:abs_frac_lx_allz} indicate that, to a first approximation, redshift dependencies
can be ignored for low- to moderate luminosity AGN. Therefore, we have carried out a study of the AGN fraction evolution as a function of
intrinsic X-ray luminosity and host galaxy stellar mass across the entire redshift range spanned by our sample.
The results are shown in the left panel of Figure~\ref{fig:abs_frac_lx_mass}. Clearly, the total stellar mass 
of the host galaxy does not appear to have any strong influence on the fraction of the optically obscured AGN 
in our sample, even if there is marginal
evidence of a slightly larger fraction of obscured AGN in the most massive galaxies. 
The right hand panel of the same figure shows again the obscured AGN fraction for different stellar
mass bins, but now plotted as a function of the specific accretion rate. 

Not surprisingly, given what we discussed in section~\ref{sec:frac_class} and on the left panel of 
Fig.~\ref{fig:abs_frac_lx_mass} itself, the specific accretion rate does not seem to be a good predictor
of the obscured AGN fraction, and definitely it is a much worse one than the X-ray luminosity itself. Indeed,
as shown by the dashed lines of the right panel of Fig.~\ref{fig:abs_frac_lx_mass}, the data are consistent 
with the simplest assumption that the fraction of (optically defined) type--2 AGN is {\it mainly} a function
of luminosity, so that neither the Eddington ratio, not the overall gravitational potential of the host galaxy 
affects in a significant way the statistical properties of the optically obscuring medium. 

\subsection{Relationship between AGN obscuration and host galaxies star formation}
We have then analysed the relationship between AGN obscuration and the star-forming properties of their
host galaxies. To avoid as much as possible the contaminating effects of the AGN itself for
the determination of star-formation rate from optical-UV SED fitting, 
we have used the FIR data collected by {\it Herschel/PACS} \citep{Poglitsch2010} 
guaranteed time, that has 
observed the entire COSMOS field as a part of the PEP-survey \citep[][{\tt www.mpe.mpg.de/Research/PEP}]{Lutz2011}. \citet{Rosario2012} have studied in detail the possible influence of AGN-driven
emission on the {\it Herschel/PACS} fluxes and stacks of X-ray selected sources in various deep 
fields. They concluded that AGN contamination is minimal, and affects only the objects with the lowest SFR.
At the depth
of the {\it Herschel-PEP} survey of the COSMOS field, however, only a relatively small fraction of the X-ray sources
in our sample are individually detected ($\sim 18$\%). We thus resort to a stacking analysis to
assess statistically the incidence of star formation in the host galaxies of our AGN.

In order to do so, we have stacked the {\it Herschel} fluxes, both at 100 and 160 $\mu$m, 
at the location of the detected counterparts of the AGN. We did so separately for objects classified as type--1
and type--2 in each of the 15 bins of the $L_{\rm X}-z$ plane where our survey is complete (see Fig.~\ref{fig:lx_z}).
We define ``detection'' whenever the stacked signal exceeds a value of three times the 
noise in at least one PEP band in a bin.
We then derive the SFR of each redshift, luminosity
and obscuration-class bin using their stacked PACS flux densities and the \citet{Chary2001} SED library. 
In the case where there are two PACS detections (at both 100$\mu$m and
160$\mu$m), we fit the PACS flux densities with the \citet{Chary2001} 
SED library leaving the normalization of each template as
a free parameter. 
If, on the other hand, the stacked flux is detected significantly in only one PACS 
band (either at 100$\mu$m or 160$\mu$m), the 
SFR is obtained simply using the scaled SED library. As it was shown by \citet{Elbaz2010,Elbaz2011}, 
scaling the FIR templates to match a monochromatic luminosity provides robust (and non-degenerate) estimates
of the total FIR luminosity, and thus of the SFR.
In both cases, SFR is given by integrating the
best SED template from 8 to 1000$\mu$m
and using the standard relation between IR luminosity and star formation rate 
of \citet{Kennicutt1998} for a \citet{Chabrier2003} IMF. 
For bins where no stacked signal is detected, we derive robust upper limits for the average SFR. 

\begin{figure}
\includegraphics[width=0.48\textwidth,clip]{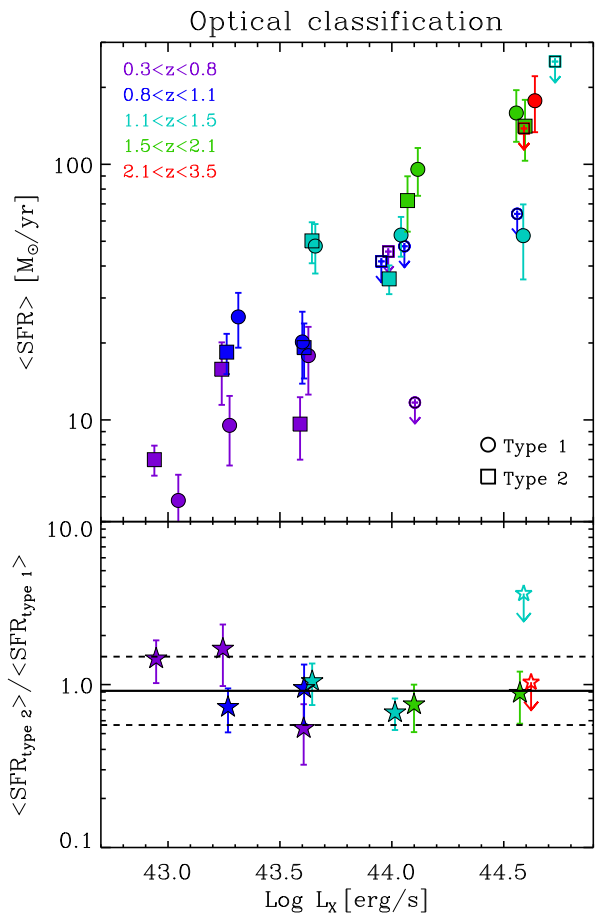}
\caption{{\it Top Panel}: The mean star formation rates for AGN in
  different luminosity and redshift bins are plotted as a function of
  AGN X-ray luminosity. Different colors mark different redshift
  interval, with the same color-code of
  Fig.~\ref{fig:abs_frac_lx_allz}. Circles represent type--1 AGN and
  squares type--2 AGN. Empty symbols with downwards-pointing arrows
  mark upper limits. {\it Bottom Panel}: The ratio of the mean star
  formation rates in obscured (type--2) and un-obscured (type--1) AGN is plotted as a
  function of their X-ray luminosity. The solid line is the best fit mean
  value, with the uncertainty marked by the dashed lines.} 
\label{fig:sfr_lx_allz}
\end{figure}

The top panel of Figure~\ref{fig:sfr_lx_allz} shows the average SFR for each of the 30 stacks we have considered, as 
a function of the average X-ray luminosity of the AGN that were included in the stacks. Different symbols distinguish 
type--1 (circles) from type--2 AGN (squares). The overall trend of the average star formation rate as a function of 
AGN X-ray luminosity mostly reflects the influence of distance on both quantities in our flux-limited sample. 
Within individual redshift intervals (points of the same color), there is little evidence of a correlation between
$L_{\rm X}$ and $\langle SFR \rangle$, a result consistent with previous investigations of SFR in AGN hosts 
from {\it Herschel} survey data \citep{Mullaney2012,Rosario2012}. This might be due to the lack
of any physical connection between SFR and accretion luminosity, or to the effect of rapid AGN
variability modulating a long term correlation \citep{Hickox2013}.

Most striking, however, is the comparison between the average star formation rate of optically obscured and un-obscured
AGN, shown in the bottom panel of Fig.~\ref{fig:sfr_lx_allz}. 
In fact, for essentially all bins in the luminosity-redshift plane for which we have at least one measure of the 
average SFR from the stacks, we see no difference whatsoever in the rate of star formation in type--1 and type--2 AGN,
despite the wide range of luminosities probed. Clearly, galaxy-wide star formation does not appear to distinguish 
between type--1 and type--2 AGN, or, to put it differently, the physical entity responsible for the optical 
classification into obscured and un-obscured AGN is not strongly affected by the overall star-formation properties
of the host galaxy.

\begin{figure*}
\includegraphics[width=0.45\textwidth,clip]{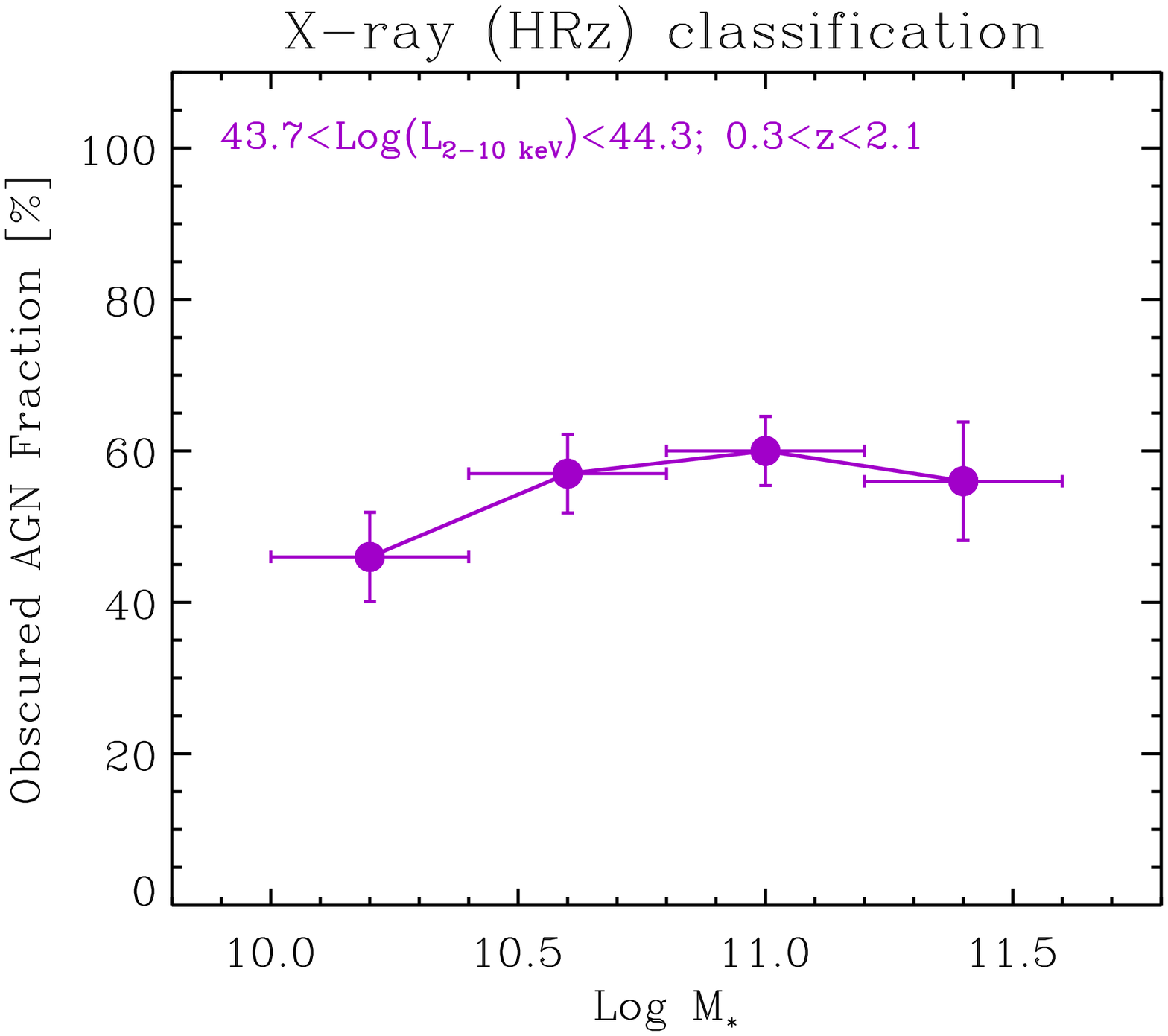}
\includegraphics[width=0.45\textwidth,clip]{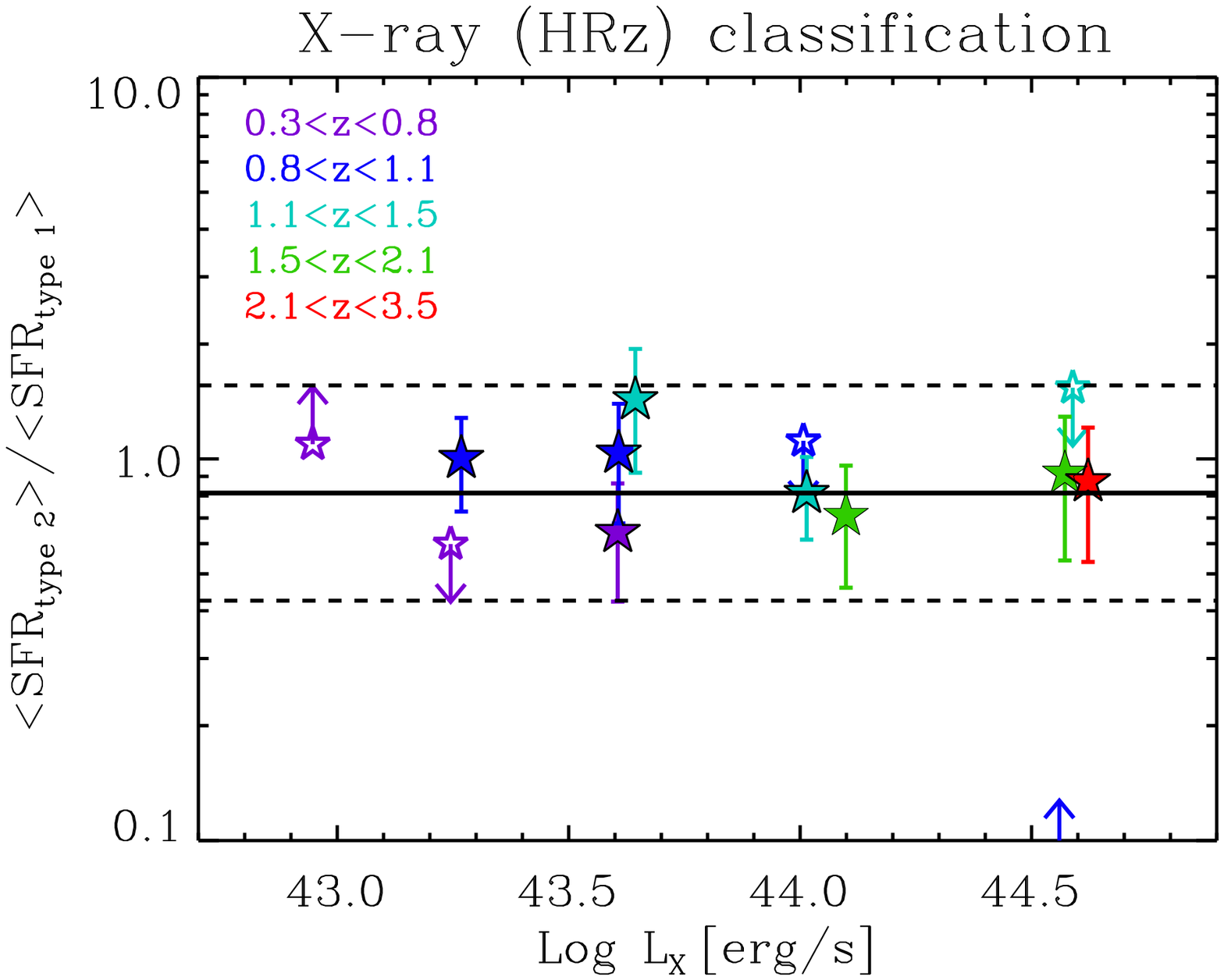}
\caption{{\it Left}: Fraction of obscured AGN, as classified from their X-ray spectra with the HRz method, 
as a function of their host galaxies stellar mass, for all AGN in the sample with $43.7<{\rm Log}L_{\rm X}<44.3$
and  $0.3<z<2.1$. {\it Right}: The ratio of the mean star
  formation rates in HRz classified as obscured (type--2) and un-obscured (type--1) AGN is plotted as a
  function of their X-ray luminosity. The solid line is the best fit mean
  value, with the uncertainty marked by the dashed lines.} 
\label{fig:host_hrz}
\end{figure*}

Despite the intriguing differences in the AGN X-ray and optical classification, the main results presented in this section
 do not change qualitatively when we look at the host galaxies of AGN classified 
on the basis of their X-ray spectral properties. 
In the left panel of figure~\ref{fig:host_hrz} 
we show the fraction of X-ray obscured AGN as a function of stellar mass of their host galaxies. In order
to improve the statistics, we restrict ourselves to the luminosity range $43.7<{\rm Log}L_{\rm X}<44.3$, 
where we can combine all 
objects within a wide redshift range ($0.3<z<2.1$), as the luminosity dependence is negligible there. Again, no strong 
dependence of the absorption fraction on $M_*$ is evident. In the right hand panel of the same figure, we see that 
the average SFR computed from {\it Herschel-PEP} stacking analysis on the X-ray obscured and
un-obscured AGN confirms our finding that the galaxy-wide SFR appears to be totally disconnected from the nuclear
X-ray obscuring medium. This is consistent with the results of large {\it Herschel} studies over similar redshifts which find no clear dependence of SFR on X-ray obscuration \citep{Shao2010,Rosario2012}.

\section{Discussion: optical vs. X-ray obscuration classification}
\label{sec:x_vs_opt}

\begin{figure}
\includegraphics[width=0.49\textwidth,clip]{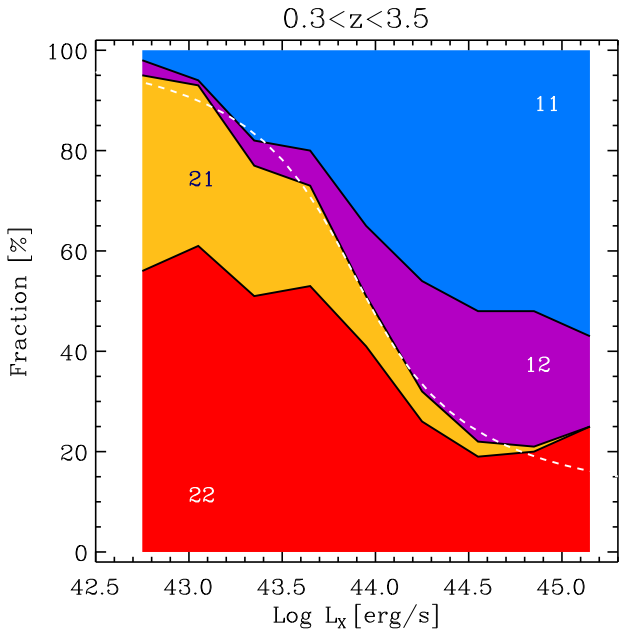}
\caption{The relative fractions of AGN classified according to our 4-ways scheme are plotted
as a function of intrinsic X-ray luminosity. Red shaded area is for type--22 AGN (both X-ray and optically obscured);
blue for type--11 (both X-ray and optically un-obscured), while the orange shaded area is for type--21 objects 
(those which are X-ray un-obscured but optically obscured) and the purple one for type--12 AGN 
(optically un-obscured but X-ray obscured).} 
\label{fig:abs_frac_lx_allz_hrz_both}
\end{figure}

As we showed in Section~\ref{sec:frac_class}, the differences in the optical and X-ray based classification of the AGN are 
strongly luminosity dependent. We can show this if we introduce a 4-ways classification of the entire sample.
Within such a scheme, we can define {\it type--11} AGN those which appear un-obscured both from their optical and X-ray 
spectra (363 in total); {\it type--22} objects are instead those which appear obscured both at X-ray and optical wavelengths (546). 
Finally, we call {\it type--12} AGN those which are optically un-obscured (showing broad lines in their optical
spectra and/or a strong, blue optical/UV continuum and a point-like morphology) but have X-ray spectra (or 
hardness ratios) consistent with a value of $N_{\rm H}>10^{21.5}$ cm$^{-2}$ (167); conversely we call {\it type--21} those
which are consistent with no X-ray obscuration, but does not show broad lines in their optical spectra, or have 
galaxy-like optical/UV Spectral Energy Distributions (234). 

Figure~\ref{fig:abs_frac_lx_allz_hrz_both} shows the distribution of these four classes as a function of 
luminosity (across the entire redshift range spanned by our sample). The $\sim$30\% AGN with ambiguous classification clearly
separate into two distinct luminosity regimes: type--21 (orange shaded area) at low- and type--12 (purple shaded
area) at high luminosities. We now discuss in greater detail the properties of these classes of objects.

\subsection{Clues on the nature of the optical/X-rays discrepant classification: stacked spectra}
\label{sec:stack}
The two sets of AGN with discordant classification call for a physical explanation. Having shown that this is most likely
not to be searched in a different property of the host galaxies, we need to investigate in more detail the 
properties of the nuclear emission on parsec and sub-parsec scales, where the X-ray and optical features 
(big blue bump and/or broad lines) used to classify the objects are produced.

First, we have looked at the average spectral shape of four separate groups of objects.
As we discussed before, there are two clearly separate regimes we need to study. At low luminosities (and,
in our flux-limited sample, at low redshift) many X-ray un-obscured AGN are optically classified as obscured 
(type--21 class). In order to sample this regime, 
we took a complete sample, selected in the redshift range $0.3<z<0.9$ and with 
$42.9<{\rm Log} L_{\rm X}<43.5$ (see the leftmost 
yellow area in Fig.~\ref{fig:lx_z}), and constructed the average X-ray and optical spectra 
for the type--21 (57 objects, 47 of which with spectroscopic redshift information) and type--22 AGN
(99 objects, 61 with spec-z), respectively. On the other hand, in order to study the properties of the 
high-luminosity AGN that are classified as obscured from the HRz method, but appear optically un-obscured, 
 we took a complete sample, selected in the redshift range $1.5<z<2.5$ and with 
${\rm Log} L_{\rm X}>44$ (rightmost yellow area in Fig.~\ref{fig:lx_z}), and constructed the average X-ray and optical spectra 
for the type--11 (109 objects, 94 with spectroscopic redshifts) and type--12 AGN (66 objects, 38 with spec-z), 
respectively. 

\begin{figure}
\includegraphics[width=0.45\textwidth,clip,angle=270,trim=-20 0 0 0]{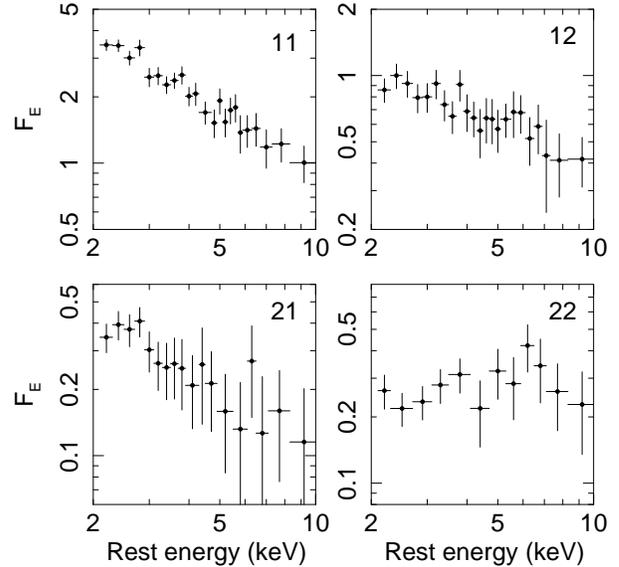}
\caption{Rest frame stacked X-ray spectra for the four classes of objects in two different regions of the 
$L_{\rm X}-z$ plane (see yellow boxes in Fig.~\ref{fig:lx_z} and section~\ref{sec:stack}). The top row
shows the stacked spectra for type--11 and type--12 AGN, respectively, 
in the $1.5<z<2.5$ redshift range and with $L_{\rm X}>10^{44}$ erg/s. The bottom row 
shows the stacked spectra for type--21 and type--22 AGN, respectively, 
in the $0.3<z<0.9$ redshift range and with $10^{42.9}<L_{\rm X}<10^{43.5}$ erg/s.} 
\label{fig:stack_x}
\end{figure}

For the X-ray rest-frame spectral stacking, we adopt the method of \citet{Iwasawa2012}, and produce rest-frame
stacked spectra in the 2-10 keV range using the {\it XMM-Newton} data, and
the four rest-frame stacked X-ray spectra are shown in Figure~\ref{fig:stack_x}.  
We also compared the optical properties of the samples by stacking all available 
optical spectra. For this latter exercise, as we combine spectroscopic observations with 
different instruments (SDSS, IMACS, VLT/VIMOS from the zCOSMOS bright and deep surveys), covering different 
wavelength ranges at different resolutions, we first smooth all spectra to the resolution of the worst spectrum 
available (set at R=300 by the zCOSMOS deep program for the high redshift type--11 and type--12 samples and at R=600 
by the zCOSMOS bright spectra for the low redshift type--21 and type--22 samples), 
 then normalize at a common rest-frame wavelength 
(2200 {\AA} for the type--11 and type--12 at high redshift, and  4500 {\AA} 
for the type--21 and type--22 at low redshift) and finally compute the mean rest-frame spectrum.

\subsubsection{The most luminous AGN: closing in on the close-in absorber}
\begin{figure}
\includegraphics[width=0.38\textwidth,clip,angle=270]{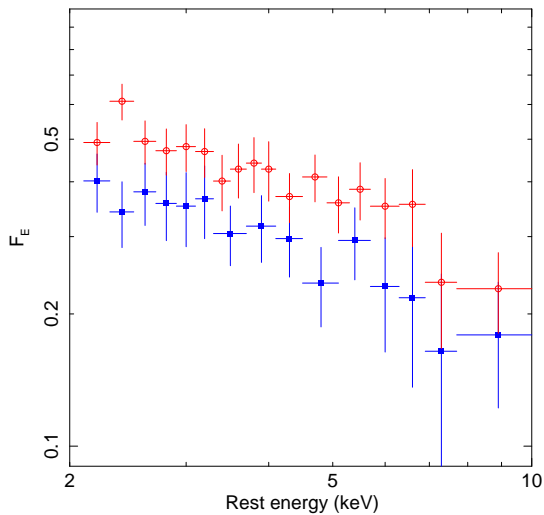}
\caption{Rest frame stacked X-ray spectra for type--12 AGN with spectroscopic redshift information (red points) and with photo-z (blue points). 
The spectroscopic sample is the same used to produce the median optical spectrum shown with a purple line in Fig.~\ref{fig:stack_spectra_all} below.
Both samples can be described by a simple power-law model fit with spectral index $\sim$1.5, or  
with an absorbed power-law model of $\Gamma_{12} \simeq 1.7$ and $N_{\rm H} \simeq 7 \times 10^{21}$ cm$^{2}$.} 
\label{fig:stack12_specz_photoz}
\end{figure}

The top row of Figure~\ref{fig:stack_x} shows the two stacked X-ray 
spectra for the two high-redshift, high-luminosity groups: type--11 (upper left) and type--12 (upper right).
As expected, all luminous AGN classified as un-obscured from both X-ray and optical analysis show an almost featureless
power-law spectrum, with slope $\Gamma_{11}=1.9 \pm 0.1$, consistent with the basic assumption of the HRz 
method. No clear evidence of emission lines is seen, possibly consistent with the so-called Iwasawa-Taniguchi effect, 
i.e. the observational fact that the equivalent width of the ubiquitous narrow Fe K$\alpha$ emission line decreases
with increasing X-ray luminosity for the AGN \citep{Iwasawa93,Bianchi2007,Chaudhary2010}. 
The average rest-frame X-ray spectrum of the type--12 AGN has a significantly shallower slope: when fitted with a simple power-law, we obtain $\Gamma_{12}=1.5 \pm 0.1$. However, an equally good
fit can be obtained with an absorbed power-law model 
with $\Gamma_{12} \simeq 1.7$ and $N_{\rm H} \simeq 7 \times 10^{21}$ cm$^{-2}$, 
suggesting that indeed there could be excess absorption in these optically classified, highly luminous, type--1 AGN.
Note that the difference between type--11 and type--12 median X-ray spectra cannot be due to the 28 type--12 objects
without spectroscopic information, as 
type--12 objects with and without spectroscopic redshift show almost identical rest-frame stacked X-ray spectra, 
as can be seen in Figure~\ref{fig:stack12_specz_photoz}.

Incidentally, the fact that no significant iron emission line is detected in these two stacked spectra
can be used to argue against a large fraction of these sources being hidden behind a Compton Thick screen
with high covering fraction, as required by the model of \citet{Mayo2013} to explain 
the observed obscured fraction vs. luminosity relation.

\begin{figure*}
\includegraphics[width=0.96\textwidth,clip]{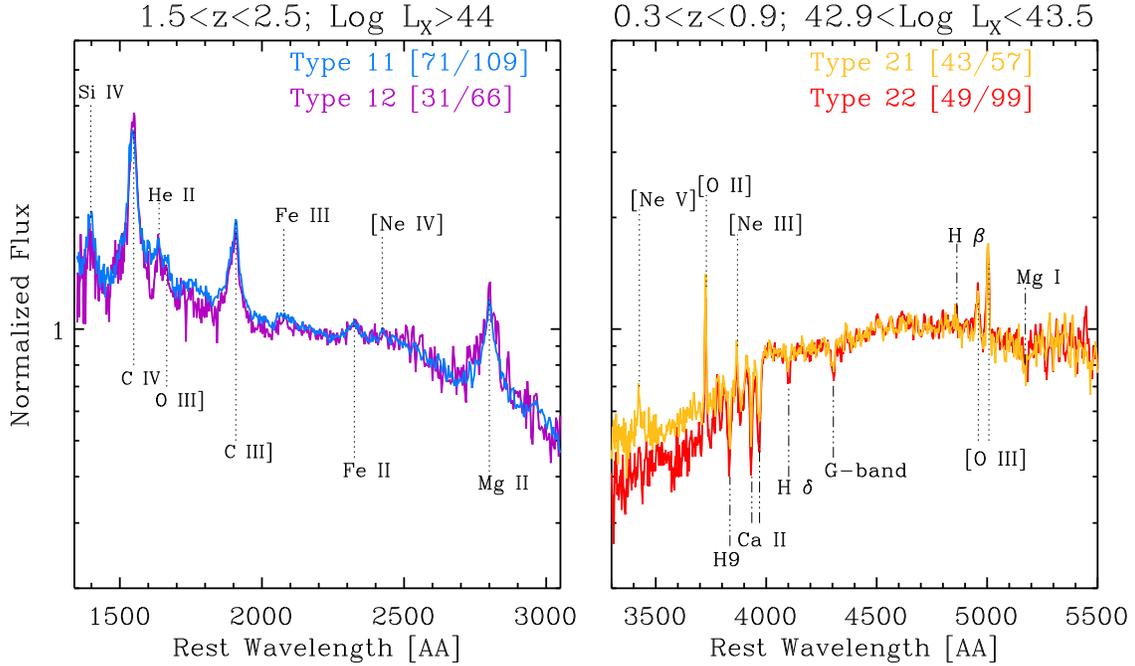}
\caption{{\it Left:} Rest frame median spectra for the two high-redshift, high-luminosity
 classes of objects, type--11 and type--12, are plotted with blue and purple lines, respectively. The most prominent
emission lines are marked. Spectra are normalized at 2200 {\AA}. 
{\it Right:} Rest frame median spectra for the two low-redshift, low-luminosity
 classes of objects, type--21 and type--22, are plotted with orange and red lines, respectively. 
The most prominent emission and absorption lines are marked. Spectra are normalized at 4500 {\AA}.} 
\label{fig:stack_spectra_all}
\end{figure*}

The left hand panel of Figure~\ref{fig:stack_spectra_all} shows the comparison between the rest-frame median 
optical spectra of type--11 and type--12 AGN.
The spectroscopic completeness of the two samples is significantly different: while 
the template spectrum for the type-11 AGN is made with 71/109 object in the complete X-ray selected sample, we have only
31/66 spectra for the type--12. Nevertheless, the stacked optical spectra of the two classes are very similar, with no 
clear evidence that the type--12 are more reddened than the type--11, and no significant
discrepancy in the equivalent width of the most prominent broad emission lines. Moreover,
the median SED derived by stacking the best fit models of all photometric observations for type-11 and type--12 AGN
in this redshift and luminosity range do not show any 
substantial difference either, as can be seen in the left panel of Figure~\ref{fig:stack_SED_all}.
This suggests that large-scale (or galaxy-wide) extinction is not the cause of the X-ray obscuration, and 
contradicts the naive expectation of AGN evolutionary models in which powerful, obscured QSOs are associated
to galactic absorption produced by the vigorous star-formation activity. While the most luminous AGN in our sample
tend to be hosted by galaxies with high star-formation rate, there is no evidence of any relationship between this
and the obscuration properties of the sources (see Figure~\ref{fig:sfr_lx_allz} and~\ref{fig:host_hrz}).

In fact, it seems more plausible to argue that the excess X-ray obscuration in a large fraction of the most
luminous AGN could be produced by dust-free gas within (or inside) the broad line region. 
In the nearby Universe, the presence of small-scale (sub-parsec)
absorbers has been inferred in a number of cases \citep{Risaliti2002,Maiolino2010}, with NGC 4151 being the prototypical example,
well known since more than thirty years \citep{Ives1976}. Evidence of
excess absorption in more luminous, distant Type-1 QSOs was already highlighted by \citet{Perola2004}, \citet{Fiore2012} 
and, for AGN in the 
{\it Chandra}-COSMOS survey, by \citet{Lanzuisi2013}. An interesting possibility, worth further exploration, is that 
type--12 AGN might be related to Broad absorption-line quasars (BALQSOs). BALQOs have indeed long been known to be
X-ray obscured \citep[see e.g.][]{Green1995,Gallagher1999,Brandt2000}, even if more recent works on X-ray selected samples 
questioned this evidence \citep{Giustini2008}. Due to the lack of a systematic analysis of the optical
spectra of broad-line AGN in the zCOSMOS and IMACS spectroscopic surveys and the inclusion of photometric type 1 AGN,
we do not have a robust estimate of the BAL fraction in our sample. Assuming they constitute about 10-15\% of the entire quasar
population \citep{Richards2003,Hewett2003}, BALQSOs may be
numerous among type--12 AGN. Indeed, a tantalizing hint of extra absorption can be seen in the stacked spectrum of type--12 AGN
in Figure~\ref{fig:stack_spectra_all}.

\subsubsection{The nature of Type--21 AGN at low luminosity}

Let us now consider the opposite end of the luminosity scale: 
among lower luminosity AGN (with $L_{\rm X} < 10^{44}$ erg/s), a substantial fraction (about 1/3) is optically classified as 
type--2, but is un-obscured in the X-rays, and we called these objects type--21 AGN.
These might well include so-called 'true' type--2 AGN \citep{PanessaBassani2002,Brightman2008,Bianchi2012a}, i.e. 
accreting black holes without a hidden broad line region. Our sample of 234 such objects is about 
one order of magnitude larger than any previous one.   

In the local Universe, objects of this class
 had been identified among Seyfert 2 galaxies by the lack of polarized broad emission lines \citep{Tran2001,Tran2003}, 
and a number of theoretical investigations have explored models for the BLR and/or for the obscuring torus which naturally
account for their disappearance at lower luminosities \citep{Nicastro2000,Nenkova2008,Elitzur2009}, even if in most
models this effect sets in at much lower luminosities and/or Eddington ratios than those probed by our sample. 
We cannot discard this
as a viable interpretation of our data, but we note that the concurrent effect of galaxy dilution, i.e. the decreasing contrast
between the nuclear AGN emission and the stellar light from the whole galaxy must play an important role, too.

A simple estimate illustrates this point \citep[see e.g.][]{Merloni2013}. Let us consider an
AGN with optical B-band luminosity given by $L_{\rm AGN,B}=\lambda L_{\rm Edd}
f_{\rm B}$, where we have introduced the Eddington ratio ($\lambda\equiv L_{\rm bol}/L_{\rm
  Edd}$)\footnote{$L_{\rm Edd}=4 \pi G M_{\rm BH} m_{\rm p} c /
\sigma_{\rm T} \simeq 1.3 \times 10^{38} (M_{\rm BH}/M_{\odot})$ ergs
s$^{-1}$ is the Eddington luminosity}, and a bolometric correction 
$f_{\rm B}\equiv L_{\rm AGN,B}/L_{\rm bol}\approx 0.1$
\citep{Richards2006SED}.  Assuming a bulge-to-black hole mass ratio of
0.001 and a bulge-to-total galactic stellar mass ratio of $(B/T)$, the
contrast between nuclear AGN continuum and host galaxy blue light is
given by:
\begin{equation}
\label{eq:agn_contrast}
\frac{L_{\rm AGN,B}}{L_{\rm
    host,B}}=\frac{\lambda}{0.1}\frac{(M_*/L_{\rm B})_{\rm
    host}}{3(M_{\odot}/L_{\odot})}(B/T)
\end{equation} 
Thus, for typical mass-to-light ratios, the AGN will become
increasingly diluted by the host stellar light in the UV-optical-IR bands 
at Eddington ratios $\lambda$ smaller than a few per cent, which is quite common among lower 
luminosity type--2 AGN in the sample (see Figure~\ref{fig:lx_m}). 

Broadly speaking, in order to understand the nature of type--21 AGN, three explanations can be put forward \citep[see e.g.][]{Bianchi2012a}, 
depending on how much do the X-ray spectra reveal the true un-absorbed nature of these objects: (i) low-luminosity AGN 
do not produce broad lines, as the physical conditions in the accretion disc and its associated outflows are not
favorable for the generation of a broad-line region \citep{Nicastro2000,Elitzur2009}; 
(ii) these weak type--1 AGN are simply
out-shined by the stellar light from the massive host galaxy they live in, especially at the relative longer wavelengths 
probed by the majority of spectrographs used to follow-up XMM-COSMOS AGN in this luminosity range, and (iii) these objects 
are indeed heavily obscured (Compton Thick, CT), 
and the soft X-ray spectra appear because of a strong scattered continuum \citep{Brightman2012a,Mayo2013}. NGC 1068 
\citep{Pier1994} would be the prototypical analog in the nearby Universe. 

In the first case we expect no difference whatsoever between the optical/NIR SED of objects with obscured or un-obscured 
X-ray spectra, while in the second case a weak nuclear continuum should emerge at shorter wavelengths. In the CT case, 
the optical spectra of type-22 and type-21 should be similar, but the type--21 should have a much stronger hot dust component
in the NIR.

The bottom row of Fig.~\ref{fig:stack_x} shows the two X-ray
spectra for the two low-redshift, low-luminosity groups: type--21 (lower left) and type--22 (lower right).
The type--22 AGN show a clear signature of obscured spectra, with an apparent curvature, and an
overall shape that, if fitted with a simple power-law, would require a very hard slope $\Gamma_{22}=0.8 \pm 0.2$. 
On the other hand, the vast majority of AGN which are HRz-classified as un-obscured in the type--21 class
must indeed not show any evidence of cold, neutral absorption: the stacked X-ray spectrum, despite the low signal to 
noise, can be fitted with a power-law of slope $\Gamma_{21}=1.8\pm 0.2$. A tantalizing excess consistent with emission 
from Iron K$\alpha$ at 6.4 is apparent (with Equivalent Width $\sim$ 300 eV), but significant at just the 2$\sigma$
level; where this the telltale of heavily obscured AGN (which have typically EW of the order of 1 keV), one could speculate that
a fraction of the order of 20\% of the type--21 AGN could harbour heavily obscured nuclei. 

Still, for the majority of the type--21 classification the fundamental question we need to answer is: why do the optical
spectra of these low-luminosity AGN, the X-ray spectra of which are by and large un-obscured, not show any evidence of 
un-obscured AGN emission?

\begin{figure*}
\includegraphics[width=0.98\textwidth,clip]{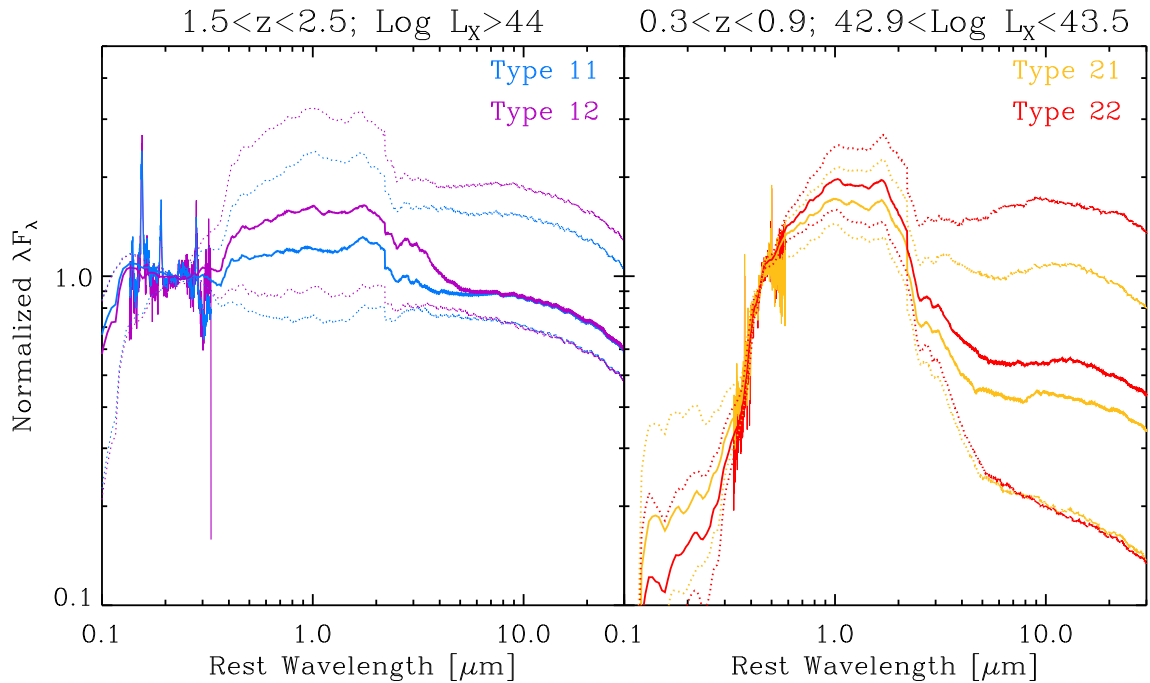}
\caption{{\it Left:} Rest-frame median SED for type--11 and type--12 AGN (blue and purple solid lines, respectively). 
The dashed lines mark the 25$^{\rm th}$ and 75$^{\rm th}$ percentiles. The median spectra for the spectroscopic samples, 
all normalized at the same rest-frame wavelength (2200 \AA)
are over-plotted. {\it Right:} Rest-frame median SED for type--21 and type--22 AGN 
(orange and red solid lines, respectively). 
The dashed lines mark the 25$^{\rm th}$ and 75$^{\rm th}$ percentiles. The median spectra for the spectroscopic samples, 
all normalized at the same rest-frame wavelength (4500 \AA)
are also over-plotted. In both panel, and for all four AGN groups, we note a substantial agreement between the SED
of the complete samples, and of the spectroscopic ones.} 
\label{fig:stack_SED_all}
\end{figure*}

\begin{figure}
\includegraphics[width=0.49\textwidth,clip]{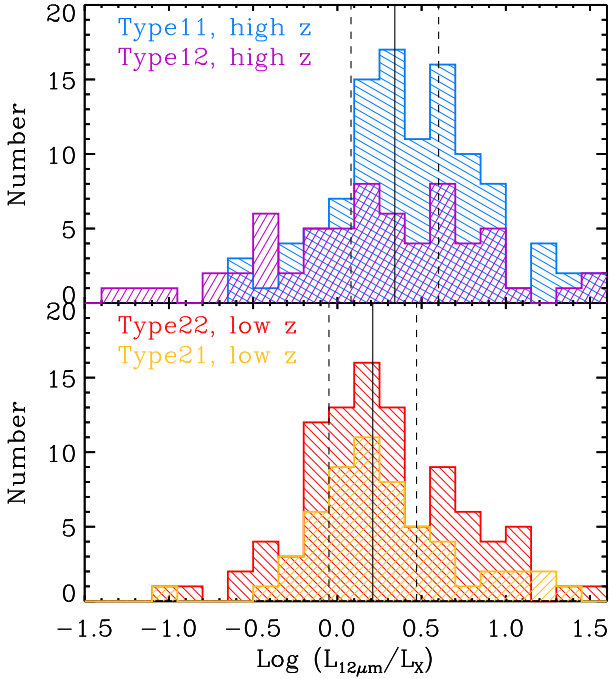}
\caption{Distribution of the ratio between MIR (rest-frame 12$\mu$m) and intrinsic X-ray luminosities for the four selected sub-samples discussed in the text. The top panel shows type--11 and type--12 AGN in the high-redshift, high-luminosity range, while the bottom shows the type--21 and type--22 AGN in the low-luminosity, low-redshift range. The vertical solid line mark the expected ratio (at the corresponding luminosity)
for un-obscured, un-contaminated 
AGN from the relation of \citet{Gandhi2009}, with the dashed lines denoting 1-$\sigma$ uncertainty. This is an indirect confirmation that our
absorption correction to the X-ray luminosity is, statistically, correct, as therefore should be our HRz classification.} 
\label{fig:lir_lx}
\end{figure}

To further examine this issue, we show in the 
right hand panel of Figure~\ref{fig:stack_spectra_all} the comparison between the 
rest-frame median spectra of type--21 and type--22 AGN.
As opposed to the case of the type--11 and type--12 median spectra, we see now two clearly galaxy-dominated spectra,  
displaying a strong stellar continuum with superimposed  absorption features seen typically in massive galaxies.
Also the global SED derived from combining the best-fit models to the entire photometric data-sets, 
shown in the right panel of Figure~\ref{fig:stack_SED_all}, are dominated by the host-galaxy stellar emission all the way
to the rest-frame NIR. 
Within the wavelength range where the signal-to-noise of the stacked spectrum is high enough, 
no strong broad emission line can be seen
(the only one expected being $H\beta$), neither in the type--22 AGN, nor in the type--21 AGN composite.
However, two differences can be seen between the two composites: 
the type--21 spectrum has significantly stronger continuum 
blue-ward of the 4000 {\AA} break, and a prominent, clearly visible [Ne V] $\lambda$3426 emission line, 
that could be the signature of the optical/UV ionizing continuum 
emerging from the redder host galaxy stellar light \citep{Gilli2010,Mignoli2013}. 
In fact, the emergence of a blue continuum shortwards of the 4000 {\AA} break could
suggest that, for the majority of type--21 AGN, galaxy dilution is the most plausible explanation.
This is also confirmed by the measured ratio between mid-IR (rest frame 12$\mu$m, measured from the overall SED fit, see 
Bongiorno et al. 2012) and intrinsic 2-10 keV luminosity, which we plot in 
Figure~\ref{fig:lir_lx} for the four relevant subclasses 
(type--11 and type--12 in the top panel; type--21 and type--22 in the bottom one).

Finally, we show in Figure~\ref{fig:lir_lx}, that the $L_{12\mu{\rm m}}/L_{\rm X}$ ratio is 
similarly distributed in type--21 and type--22 objects, and, as shown already in Fig.~\ref{fig:lir_lx_all}, consistent with the  
expected ratio for a pure, un-obscured AGN, based on the best-fit
relation of \citet{Gandhi2009} (solid vertical line).
Thus, if the MIR luminosity is dominated by the AGN (as we assumed in our SED modelling, see Bongiorno et al. 2012), 
then our absorption correction to the X-ray luminosity must be by and large correct, as we discussed in 
section~\ref{sec:rest_frame_sel}. 

Further detailed studies of individual  type--21 
sources will be necessary to resolve the conundrum, including rest-frame UV and NIR spectroscopy, polarization and radio 
observations.

\section{Conclusions}
\label{sec:conclusions}
In this work, we have looked in great detail at the obscuration properties of a complete sample of
AGN selected by {\it XMM-Newton} in the COSMOS field. The unique multi-wavelength coverage of the 2 deg$^2$ 
of the XMM-COSMOS area allows a full identification of the X-ray sources counterparts, and an extraordinary 
redshift completeness, by means of both spectroscopic observations and well-calibrated photometric redshift analysis.
This, in turn, enables us to perform a basic K-correction of the X-ray spectra for all objects for which the number of 
X-ray counts is insufficient for a thorough spectral analysis. We could therefore select a large (1310 AGN) 
sub-sample of the XMM-COSMOS AGN above a fixed {\it rest-frame} 2-10 keV 
intrinsic flux (equal to $2\times 10^{-15}$ erg/s/cm$^2$), in the redshift range $0.3<z<3.5$,
conveniently mitigating the bias caused by the degeneracy between redshift and neutral absorption column density.

We classify {\it all} the AGN in our sample as obscured or un-obscured on the basis of either (i)
the optical spectral properties and/or the overall spectral energy
distribution and optical morphology or (ii) the shape of the X-ray spectrum.

To our knowledge, this is the largest systematic study of the difference between type--1 and type--2 AGN classifications based
on X-ray vs. optical spectral energy distributions \citep[see e.g.][for a comparison]{Corral2011,Malizia2012}. 
Our ability to provide optical and X-ray classifications for a large, completely identified sample of X-ray selected AGN
uncovers the complexity of the obscuring medium in the nuclear region of active galaxies. Our main conclusions are the following:

\begin{itemize}

\item{By contrasting the cumulative distribution of the inferred column density for 
optically obscured and un-obscured AGN, we found that the most consistent classification divide occurs 
at ${\rm Log}\, N_{\rm H} \simeq 21.5$, corresponding to an optical reddening of $E(B-V)=0.57$, for standard dust-to-gas ratio.
This is not too surprising, as already above $E(B-V)\approx 0.3$ the optical ($0.3-1 \mu m$) slope of a typical QSO would become 
red \citep{Bongiorno2012,Hao2013}. Nevertheless, such 'optimized' classification still fails for about 30\% of the entire sample, for
which the optical- and X-ray-based classifications give contrasting results. These 'unclassifiable' AGN clearly separate into two well
distinct classes according to their nuclear luminosity (or redshift: in our flux-limited sample we are not able to fully disentangle
these two possibilities).}

\item{By studying the stacked X-ray and optical spectra of both the low-luminosity, low-redshift type--21 AGN we conclude that their 
optical classification as obscured, despite the apparently un-obscured X-ray spectra, is most likely due to host galaxy dilution, 
even if we cannot rule out that a small minority of these objects are CT AGN or true ``naked'' type--2.}

\item{The high-luminosity, high-redshift type--12 objects, instead, appear to be truly X-ray obscured, despite their optical SED typical of type--1 QSOs, and we argue that such excess 
absorption could be produced by dust-free gas within (or inside) the broad line region.}

\item{The simplest version
of the classical ``unification by orientation'' scheme needs to be modified to account for the strong luminosity-dependence
of the obscured AGN fraction. Radiative coupling between the central AGN and the source of obscuration must be the critical
ingredient of any model, with the physics of dust sublimation probably playing a fundamental role. We cannot completely rule out
the possibility that a substantial fraction of the AGN in our sample have more complex spectra than what we have assumed. If they
are in fact dominated by almost complete covering from Compton Thick material, this could, at least partially, contribute to
artificially enhance the dependence of the obscured AGN fraction on luminosity. The discrepancy in the literature between 
the statistics of obscured AGN fractions in X-ray, optical, radio and mid-IR selected samples \citep{Lawrence2010} 
warrants further exploration of this issue. Here we simply note that, for the bulk of our sample,
the IR-to-X-rays luminosity ratio is consistent with the values of a local sample observed at high angular resolution 
\citep{Gandhi2009}, suggesting that our estimated intrinsic X-ray luminosities are not biased low.} 

\item{As far as the redshift evolution of the obscured AGN fraction is concerned, we clearly do not detect any significant evolution
in the optically obscured AGN fraction at low-to-moderate luminosities 
(see the right panels of figures~\ref{fig:abs_frac_lx_allz} and ~\ref{fig:abs_frac_lx_allz_hrz}), 
while obscured AGN (both X-ray and optically classified) at high luminosity
are more common at high redshift. Hasinger (2008) had compiled a similarly sized sample of AGN from a number of different 
surveys, and studied the evolution of the obscured fraction, defined in a hybrid way (from optical spectra, when available, and 
from X-ray hardness ratio otherwise). He found an increase with redshift of the fraction of obscured AGN, but, as remarked by
\citet{Gilli2010b}, the lack of a proper K-correction in \citet{Hasinger2008} induces a severe redshift-$N_{\rm H}$ bias (see our 
discussion in section~\ref{sec:sample} above), whereby an intrinsically constant fraction of obscured AGN as a function of redshift
would have shown strong positive evolution. Having corrected for such an effect, 
albeit in a rudimentary way, due to the poor photon statistics for most
of our sources, we can conclude that the most significant redshift evolution is displayed only by the most luminous AGN.}

\item{All measurable host galaxy properties in our sample do not show any relationship with the obscured/un-obscured
classification. Stellar masses can be measured quite accurately for the XMM-COSMOS AGN, as previously demonstrated by 
\citet{Merloni2010} and \citet{Bongiorno2012}, and shown by the characteristic shape of the 4000 {\AA} break in the median
SED of all four exemplary classes of AGN we discussed in section~\ref{sec:stack}. When controlling for the redshift and/or the
luminosity of the sources, we do not see any dependence of the obscured fraction on the host galaxy stellar mass 
(Figure~\ref{fig:abs_frac_lx_mass} and \ref{fig:host_hrz}).}

\item{Even more striking is the lack of any significant difference between the average star-formation rate of obscured and un-obscured 
AGN, as probed by the stacked {\it Herschel} fluxes at the position of the sources in the sample (Figure~\ref{fig:sfr_lx_allz}). 
This is remarkable, as the sample
spans a wide range of redshift, nuclear luminosity, and average SFR. In \citet{Bongiorno2012} we have measured indirectly the AGN
duty cycle by measuring the fraction of AGN in a complete parent sample of field galaxies as a function of nuclear luminosity,
host stellar mass and specific accretion rate. There we found a universal functional form that describes the probability of a 
galaxy to host an AGN as a function of the specific accretion rate, the normalization of which increases with redshift 
in steps with the increase of the specific star-formation rate. This led to the suggestion that, 
in a statistical sense, AGN activity and star formation may be globally correlated, but that there is little physical connection in 
each individual source between the gas accreting onto the SMBH and the material out of which stars form throughout the galaxy.
Here we confirm this general view, by demonstrating the lack of any connection between the matter obscuring the accreting black holes
(on pc or sub-pc scales) and the galaxy-wide SFR, consistent with the conclusions of large comprehensive Herschel studies \citep{Rosario2012}.}

\end{itemize}

In closing, we note that, despite the large size of the complete sample of X-ray selected AGN
studied here, we are still limited by small number statistics once we attempt to unravel the
true nature of the relations between obscuration ($N_{\rm H}$), luminosity, redshift, host galaxy 
stellar mass and star-formation rate. Arguably, only the next generation of massive AGN surveys enabled by the {\it eROSITA} all-sky
X-ray survey \citep{Merloni2012,Kolodzig2013}, with its expected harvest of hundreds of thousands of 
AGN, will enable the necessary leap forward, provided extensive multi-wavelength coverage can be obtained over
very wide areas of the sky. 

\section*{Acknowledgments}
We thank K. Nandra, M. Brightman, J. Buchner, M. Elitzur for useful discussions, and the referee,
Andy Lawrence, for his insightful comments and suggestions.
AM and MS acknowledge financial support from the DFG cluster of excellence
``Origin and structure of the universe'' ({\tt www.universe-cluster.de}).
AB is supported by the INAF fellowship program.
KI thanks support from Spanish Ministerio de Ciencia e Innovaci\'on (MICINN) through the grant AYA2010-21782-C03-01.
AC acknowledges financial contribution from the agreement ASI-INAF I/009/10/0 and INAF-PRIN 2011.
The HST COSMOS Treasury program was supported through NASA grant
HST-GO-09822. This work is mainly based on observations obtained with
{\it XMM-Newton}, an ESA Science Mission with instruments and contributions
directly funded by ESA Member States and the USA (NASA), and with the
European Southern Observatory under Large Program 175.A-0839,
Chile. In Germany, the XMM-Newton project is supported by the
Bundesministerium f{\"u}r Wirtschaft und Technologie/Deutsches Zentrum
f{\"u}r	Luft und Raumfahrt (BMWI/DLR, FKZ 50 OX 0001), the Max-Planck
Society, and the Heidenhain-Stiftung. Herschel is an ESA space observatory with science instruments provided by European-led Principal Investigator consortia and with important participation from NASA. PACS has been developed by a consortium of institutes led by 
MPE (Germany) and including UVIE (Austria); KU Leuven, CSL, IMEC (Belgium); CEA, LAM (France); MPIA (Germany); 
INAF-IFSI/OAA/OAP/OAT, LENS, SISSA (Italy); IAC (Spain). This development has been supported by the funding agencies 
BMVIT (Austria), ESA-PRODEX (Belgium), CEA/CNES (France), DLR (Germany), ASI/INAF (Italy), and CICYT/MCYT (Spain).
We acknowledge the use of the TOPCAT software ({\tt http://www.starlink.ac.uk/topcat/}).
 We gratefully acknowledge the contribution of the entire COSMOS
collaboration; more information on the COSMOS survey is available at
http://www.astro.caltech.edu/cosmos.  


\footnotesize{
\bibliographystyle{mn2e_mod}
\bibliography{angi}
}

\bsp
\label{lastpage}

\end{document}